\documentclass[%
 reprint,
 amsmath,amssymb,
 aps,
]{revtex4-2}

\usepackage{float}
\makeatletter
\let\newfloat\newfloat@ltx
\makeatother

\usepackage{graphicx}
\usepackage{dcolumn}
\usepackage{bm}
\usepackage{graphicx}
\usepackage{amsmath}
\usepackage{amssymb}
\usepackage{subfig}
\usepackage{qcircuit}
\usepackage{algorithm}
\usepackage{algpseudocode}
\usepackage{cprotect}
\usepackage{lipsum}
\usepackage{booktabs}
\usepackage{array}
\usepackage{comment}
\usepackage{amsmath, amsthm, amssymb}
\usepackage{braket}
\usepackage{bm}
\usepackage{physics}
\usepackage{graphics}
\usepackage[dvipsnames]{xcolor}
\usepackage{enumitem}
\usepackage{mathtools}
\usepackage{multirow}

\theoremstyle{definition}

\newcommand{\q}{\bm{q}}
\newcommand{\n}{\hat{\bm{n}}}
\newcommand{\z}{\bm{\varsigma}}

\newcommand{\sss}{\bm{\sigma}}

\newcommand{\A}{\mathcal{A}}

\newcommand{\trt}{\mathcal{T}}
\newcommand{\rr}{\mathcal{R}}

\newcommand{\tet}{\bm{\theta}}

\usepackage[makeroom]{cancel}

\makeatletter
\@ifundefined{sf@counterlist}{\newcommand{\sf@counterlist}{}}{}
\makeatother

\makeatletter
\renewcommand{\fnum@algorithm}{\fname@algorithm~\thealgorithm}
\makeatother

\usepackage[colorlinks=true, allcolors=blue]{hyperref}

\usepackage{cleveref}

\begin{document}

\title{Two-Gate Extensions of Free Axis and Free Quaternion Selection for Sequential Optimization of Parameterized Quantum Circuits}

\author{Joona V. Pankkonen}
\email{joona.pankkonen@aalto.fi}
 \affiliation{%
 Micro and Quantum Systems group, Department of Electronics and Nanoengineering,\\Aalto University, Finland
}%

\date{\today}

\begin{abstract}
We propose two-gate extensions of the sequential single-qubit optimizers, Free Axis Selection (Fraxis) and Free Quaternion Selection (FQS), termed Two-Gate Fraxis (TGF) and Two-Gate FQS (TGFQS), respectively. In contrast to Fraxis and FQS, which update one single-qubit gate at a time via quadratic local cost function and matrix diagonalization, TGF and TGFQS optimize two parameterized single-qubit gates simultaneously by constructing an exact quartic local cost function and optimizing it using classical optimizers. We further investigate how different gate pairing strategies affect optimization performance. Using numerical experiments on spin Hamiltonians, molecular Hamiltonians, and quantum state preparation tasks, we find that TGF and TGFQS frequently achieve a lower final relative error to the ground state energy or infidelity than their single-gate counterparts. We observe that the random and half-shifted gate pairing strategies for TGF and TGFQS perform best in many of the tested settings. In the additional finite-shot experiments on Fermi--Hubbard and transverse-field Ising model Hamiltonians, the best gate pairing strategies retain their advantage across the tested shot counts in shallow circuits. These improvements come at the cost of increased circuit evaluations per gate update, highlighting a trade-off between the power of local optimization and measurement overhead.
\end{abstract}

\maketitle


\section{Introduction}

Variational quantum algorithms (VQAs)~\cite{wecker2015progress, cerezo2021variational, bharti2022noisy, peruzzo2014variational} have emerged as a leading framework for leveraging the noisy intermediate-scale quantum (NISQ)~\cite{preskill_nisq} devices for near-term quantum computation. Parameterized quantum circuits (PQCs) provide a practical implementation framework for VQAs on current NISQ devices~\cite{benedetti2019parameterized}. They are usually composed of adjustable single-qubit parameterized and fixed two-qubit gates, such as CNOT and controlled-Z gates. PQCs are optimized in a hybrid quantum-classical optimization loop, where the measurement outcomes from the quantum device are used by a classical optimizer to update the circuit parameters~\cite{cerezo2021variational, benedetti2019parameterized}. PQCs and VQAs have already been applied to different domains such as quantum chemistry~\cite{guo2024experimental, singh2023benchmarking, bauer2020quantum, li2019variational, ma2023multiscale}, combinatorial optimization~\cite{combinatorial_opt_VQA_ref, combinatorial_VQA_ref2, combinatorial_VQA_ref3}, and machine learning~\cite{schuld2020circuit, ding2024scalable, cong2019quantum}.

Several approaches have been proposed to optimize PQCs. Gradient-based optimizers like Adam~\cite{kingma2014adam}, stochastic gradient descent (SGD)~\cite{amari1993backpropagation}, Adagrad~\cite{duchi2011adaptive}, or others~\cite{spall1998overview, dozat2016incorporating}, can be used via the parameter-shift rule~\cite{li2017hybrid, mitarai2018quantum, schuld2019evaluating}. First, the gradients are computed on quantum hardware and are then used to classically update the parameters of the PQC. Gradient-free~\cite{nelder1965simplex, powell1998direct} and hardware-specific quantum optimization methods such as quantum natural gradient (QNG)~\cite{stokes2020quantum, koczor2022quantum}, QN-SPSA~\cite{gacon2021simultaneous}, NFT~\cite{nakanishi2020sequential}, Rotosolve and Rotoselect~\cite{Ostaszewski_2021}, Free Axis Selection (Fraxis)~\cite{fraxis}, Free Quaternion Selection (FQS)~\cite{fqs}, and their hybrids~\cite{pankkonen2025improving, pankkonen2025enhancing} have been developed. Moreover, the FQS has recently been extended to optimize parameterized controlled gates~\cite{controlled_fqs}.

The optimization of PQCs faces significant challenges, including barren plateaus, where the gradients vanish exponentially as the number of qubits increases. Barren plateaus arise in various ways, the main one being the curse of dimensionality, where the dimension of the Hilbert space grows exponentially with the number of qubits, making optimization significantly harder. Other factors contributing to barren plateaus include quantum hardware noise~\cite{noise_induced_barren_plateau_vqa_ref, noise_induced_barren_plateau_vqa_ref2, noise_induced_barren_plateau_vqa_ref3}, entanglement~\cite{entanglement_induced_BP_ref1}, poor parameter initialization~\cite{kashif2024alleviating}, and ansatz expressibility~\cite{barren_plateau_mcClean_ref, expressibility_induced_BP_ref, barren_plateau_ansatz_expressibility}. Although BPs pose a significant problem in PQC optimization, several methods have been developed to alleviate BPs, which include small dynamical Lie algebras~\cite{larocca2025barren}, variable structure ansatzes e.g., ADAPT-VQEs~\cite{adapt_vqe_ref1, adapt_vqe_ref3}, and various parameter initialization strategies: Gaussian~\cite{gaussian_init}, Floquet~\cite{floquet_init}, reduced domain~\cite{reduced_domain_init}, and Gaussian mixture model~\cite{BP_gaussian_mixture_model_ref}. In addition, shallow PQCs of depth $\mathcal{O}(\log n)$ with a local cost function have been shown to alleviate BPs~\cite{cerezo2021cost}. A related observation has also been reported for the gradient-free optimizer FQS~\cite{fqs}. Another major obstacle in PQC optimization is the rapidly growing number of measurements to estimate the observables of the cost function as the number of qubits increases~\cite{VQE_practices_and_methods}. This is alleviated by methods such as unitary partitioning~\cite{Zhao2019Measurement, izmaylov2019unitary}, efficient measurement strategies~\cite{huggins2021efficient, hamamura2020efficient}, and other methods~\cite{jena2019pauli, gokhale2020n, arrasmith2020operator}. Additionally, the optimization of the PQC parameters becomes an NP-hard problem as the size of PQC grows~\cite{bittel2021training}.

We propose two-gate extensions of the sequential optimizers Fraxis and FQS, termed Two-Gate Fraxis (TGF) and Two-Gate FQS (TGFQS), respectively, in which two parameterized single-qubit gates are optimized simultaneously. The resulting local optimization problem is quartic, in contrast to the quadratic local optimization problems for Fraxis and FQS, and therefore requires constrained classical minimization rather than matrix diagonalization alone. Importantly, the two optimized gates do not need to be adjacent to each other but can be chosen anywhere in the PQC. We show that the choice of gate pairing strategy affects the convergence behavior for TGF and TGFQS. Related multi-parameter update ideas have been explored previously, but not in this form for sequential optimizers Fraxis and FQS. There exist methods to optimize multiple gates simultaneously, such as Jacobi diagonalization in Ref.~\cite{jacobi_diagonalization} (an extension of NFT), which constructs the cost function for $m$ circuit parameters for $m$ parameterized gates (in Ref.~\cite{jacobi_diagonalization}, $R_Y$ gates were used) using $3^m$ Fourier quadrature points and optimizes the cost function in Jacobi fashion. Afterward, the cost function can be optimized classically. A random gate pairing, as well as choosing gate pairs that act on the same qubit or are placed in the same layer, was used in Ref.~\cite{jacobi_diagonalization}. In contrast to such approaches, our construction is specifically tailored for Fraxis and FQS style parameterization of single-qubit gates and yields the exact quartic local cost function for simultaneous optimization of two arbitrary single-qubit gates within the sequential optimization setting. The scope of this work is to extend Fraxis and FQS to two-gate local updates and to study how gate pairing strategies affect sequential optimization performance. Moreover, this work is situated within the family of sequential optimizers such as Rotosolve (NFT), Fraxis, FQS, and the controlled gate extensions of FQS. The present work focuses on extending sequential local optimization from single-gate to two-gate updates, but the scaling of trainability, as well as the mitigation of barren plateaus, are left for future work.

The remainder of this paper is organized as follows. First, in Sec.~\ref{Methods_section}, we go through the optimization of PQCs with sequential single-qubit gate optimizers Fraxis and FQS. Then, we present the two-gate optimization methods, TGF and TGFQS, that extend the Fraxis and FQS to two-gate optimizers, respectively. In Sec.~\ref{results_section}, we provide numerical experiments for finding the ground state for the Fermi--Hubbard model and transverse-field Ising model, as well as for lithium hydride (LiH) and beryllium hydride (BeH$_2$) molecular Hamiltonians. Finally, we test TGF and TGFQS on a quantum state preparation task by maximizing fidelity to randomly sampled target states. In Sec.~\ref{conclusion_section}, we conclude our work and discuss extensions to this work and further research.

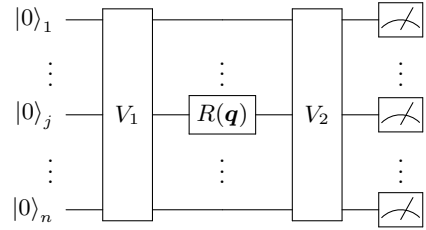
\begin{figure}
\vspace{0.1cm}
\centering
\[
\Qcircuit @C=1.5em @R=1.2em {
\lstick{\ket{0}_1} & \multigate{4}{V_1} & \qw      & \multigate{4}{V_2} & \meter  \\
\lstick{\vdots}                  &        & \vdots &          & \vdots  \\
\lstick{\ket{0}_j}  & \ghost{V_1}        & \gate{R(\q)} & \ghost{V_2}    & \meter   \\
\lstick{\vdots}                &       &  \vdots &        & \vdots  \\
\lstick{\ket{0}_n} & \ghost{V_1}        & \qw      & \ghost{V_2}        & \meter  
}
\]
\caption{Single-qubit gate tomography for sequential optimization with Fraxis and FQS.}
\label{single_qubit_gate_circuit_tomography_figure}
\end{figure}

\section{Methods} \label{Methods_section}

\subsection{Overview of Fraxis and FQS optimizers}

Sequential optimizers Rotosolve, NFT, Fraxis, and FQS update a single parameterized gate in a PQC while keeping all remaining gates fixed. Whereas Rotosolve and NFT optimize only the rotation angle $\theta$ of the gate, Fraxis and FQS optimize the single-qubit gates analytically using measurements by replacing the optimizable gate with different unitaries. After that, the optimal gate is obtained from the eigenvector of a real symmetric matrix constructed from circuit evaluations.

A general parameterized single-qubit gate $R_{\n} (\theta)$ can be written as

\begin{equation}
    R_{\n} (\theta) = e^{-i \frac{\theta}{2} \n \cdot \bm{\sigma} } = \cos\left( \tfrac{\theta}{2}\right)I - i\sin\left( \tfrac{\theta}{2}\right) \n \cdot \sss,
\end{equation}\\
where $\sss = (X, Y, Z)$ denotes the Pauli vector, $\n = (n_x, n_y, n_z) \in \mathbb{R}^3$ is the unit rotation axis, and $\theta$ is the rotation angle.  In the sequential optimization of parameterized single-qubit gates, the circuit is decomposed around a target gate acting on the $j$-th qubit, so that it can be written as

\begin{equation}
    U = V_2 R_{\n}' (\theta) V_1,
\end{equation}\\
where the single-qubit gate $R_{\n} (\theta)$ is embedded in the full $n$-qubit Hilbert space by tensoring it with identity operators. The operators $V_1$ and $V_2$ denote the parts of the circuit before and after the parameterized single-qubit gate, respectively. The parameterized single-qubit gate that acts on the $j$-th qubit in the PQC is preceded by $j-1$ identity operators and followed by $n-j$ identity operators. That is, we express the parameterized single-qubit gate as 
\begin{equation}
    R_{\n}' (\theta) = I^{\otimes (j-1)} \otimes R_{\n} (\theta) \otimes I^{\otimes (n-j)}.
\end{equation}
This ensures the correct dimensionality when the single-qubit gate acts on the $j$-th qubit for an $n$-qubit system. The cost function, defined as the expectation value of a Hermitian observable $M$, can then be written as 

\begin{equation}
    \expval{M} = \Tr\bigl[M' R_{\n}' (\theta) \rho' R_{\n}' (\theta)^\dagger \bigr].
\end{equation}\\
where $M' = V_2^\dagger M V_2$ and $\rho' = V_1 \rho_0 V_1^\dagger$. This representation isolates the dependence of the cost function on a parameterized single-qubit gate, while the remaining gates are absorbed into the transformed observable $M'$ and state $\rho'$, respectively. Here, the state $\rho_0$ denotes the initial state of the PQC that we set to $\rho_0 = \ketbra{\bm{0}}{\bm{0}}$, where $\ket{\bm{0}} \equiv \ket{0}^{\otimes n}$ and $M'$ is a $2^n \times 2^n$ Hermitian operator. Fig.~\ref{single_qubit_gate_circuit_tomography_figure} illustrates this single-qubit gate tomography viewpoint.

In Fraxis, the rotation angle $\theta$ of the gate is set to $\theta = \pi$. With this restriction, the local cost function becomes quadratic in terms of the rotation axis $\n$~\cite{fraxis}

\begin{equation}
    \expval{M}_{\n} = \n^T \Tilde{S} \n,
\end{equation}\\
where $\Tilde{S} = (S_{ij})$ is a symmetric and real $3\times 3$ matrix. The formation of the matrix $\Tilde{S}$ requires 6 circuit evaluations in total. The optimal axis $\n^*$ is obtained by computing the eigenvalues and eigenvectors of $\Tilde{S}$ and selecting the eigenvector that corresponds to the lowest eigenvalue of $\Tilde{S}$.

In FQS, this idea was extended to the fully parameterized single-qubit gate $R_{\n} (\theta)$ by encoding the rotation axis and angle to a unit quaternion $\q \in \mathbb{R}^4$. With an extended Pauli basis $\z = (I, -iX, -iY, -iZ)$ the rotation gate can be written as 

\begin{equation}
    R(\q) = \q \cdot \z.
\end{equation}\\
The local cost function in terms of the parameterized single-qubit gate then obtains the following form~\cite{fqs}

\begin{equation}
    \expval{M}_{\q} = \q^T S \q,
\end{equation}\\
where $S = (S_{\mu \nu})$ is the $4\times 4$ real and symmetric matrix whose components are given by

\begin{equation}
    S_{\mu \nu} = \frac{1}{2} \Tr\Bigl[ \left({\varsigma_\mu '}^\dagger M' \varsigma_\nu ' + {\varsigma_\nu '}^\dagger M' \varsigma_\mu' \right)\rho' \Bigr],
\end{equation}\\
where we have defined $\varsigma_\mu ' = I^{\otimes (j-1)} \otimes \varsigma_\mu \otimes I^{\otimes (n-j)} $. The FQS requires 10 circuit evaluations to construct the matrix $S$. Since the local cost function $\expval{M}_{\q}$ is quadratic in $\q$, it can be minimized (maximized) by solving the eigenvalues and eigenvectors and then selecting the eigenvector corresponding to the lowest (largest) eigenvalue. Thus, FQS optimizes the entire $SU(2)$ gate in one local optimization step. Furthermore, FQS generalizes the known sequential single-qubit gate optimizers Rotosolve, Rotoselect, and Fraxis as they are special cases of FQS~\cite{fqs}. The Fraxis can be expressed in terms of quaternions by setting the first component of the unit quaternion to zero, i.e., $\q = (0, \n)$.

\subsection{Two-Gate Optimization with Fraxis and FQS}

In this section, we further develop the Fraxis and FQS optimizers, which optimize two parameterized single-qubit gates simultaneously rather than one parameterized single-qubit gate. The single-gate optimization with Fraxis and FQS yields a quadratic local cost function, whereas two-gate updates yield a quartic local cost function. Therefore, the local optimum is no longer obtainable by diagonalization alone and must instead be found using constrained classical optimization. We fix all gates except $R_d$ and $R_k$, corresponding to the $d$-th and $k$-th parameterized single-qubit gates, respectively. We assume that $d\neq k$. The local cost function with respect to these gates is then expressed as

\begin{equation}\label{two_gate_expression_eq}
    \expval{M}_{d, k} = \Tr\left(M' R_d' V R_k' \rho' R_k'^\dagger V^\dagger R_d'^\dagger \right).
\end{equation}\\

\begin{figure}
\vspace{0.1cm}
\centering
\[
\Qcircuit @C=1.5em @R=1.2em {
\lstick{\ket{0}_1} & \multigate{6}{V_1} & \qw      & \multigate{6}{V} & \qw &   \multigate{6}{V_2} &\meter  \\
\lstick{\vdots}                  &        & \vdots &          & \vdots & & \vdots   \\
\lstick{\ket{0}_j}  & \ghost{V_1}        & \gate{R_d} & \ghost{V} & \qw & \ghost{V_2}   & \meter   \\
\lstick{\vdots}                &       &  \vdots &        & \vdots & & \vdots \\
\lstick{\ket{0}_{j'}} &   \ghost{V_1}    & \qw      &    \ghost{V}     & \gate{R_k} & \ghost{V_2} &  \meter  \\
\lstick{\vdots}                  &        & \vdots &          & \vdots & & \vdots  \\
\lstick{\ket{0}_n} & \ghost{V_1}        & \qw      & \ghost{V}     & \qw &   \ghost{V_2} & \meter  
}
\]
\caption{Tomography for two parameterized single-qubit gates for two-gate optimization with TGF and TGFQS.}
\label{two_single_qubit_gate_circuit_tomography_figure}
\end{figure}
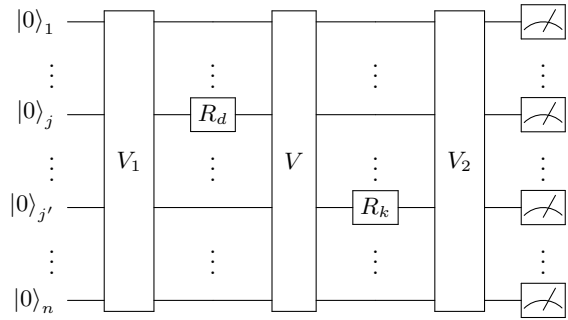

Here, the subscripts $d$ and $k$ of the cost function indicate the indices of two parameterized single-qubit gates that are optimized simultaneously, and $V$ denotes the unitary block between $R_d'$ and $R_k'$. In the special case where $R_d'$ and $R_k'$ are consecutive unitaries, the intermediate unitary $V$ reduces to the identity. We illustrate the general tomography for single-qubit gates $R_d$ and $R_k$ in Fig.~\ref{two_single_qubit_gate_circuit_tomography_figure}. Parameterized single-qubit gates $R_d$ and $R_k$ can act on the same or different qubits.

For the two-gate FQS (TGFQS) update, each parameterized single-qubit gate is represented in the extended Pauli basis as follows

\begin{equation}
    R_d' = \sum_{\mu=0}^3 q_{d,\mu} \varsigma_{d,\mu}', \quad R_k' = \sum_{\alpha=0}^3 q_{k,\alpha} \varsigma_{k,\alpha}',  
\end{equation}
where $\varsigma_0 = I$ and $\varsigma_j = -i\sigma_j$ for $j \in \{1,2,3\}$, and $\q_d, \q_k \in \mathbb{R}^4 $ are unit quaternions. By substituting these expansions into the local function in Eq.~(\ref{two_gate_expression_eq}), we obtain a quartic polynomial in terms of quaternion components

\begin{equation}
    \expval{M}_{\q_d, \q_k} = \sum_{\mu, \nu, \alpha, \beta= 0}^3 q_{d,\mu} q_{d,\nu} q_{k,\alpha} q_{k,\beta} \A_{\mu \nu \alpha \beta},
\end{equation}
with
\begin{equation}
    \A_{\mu \nu \alpha \beta} \coloneqq \Tr\left(M'\varsigma_{d,\mu}' V \varsigma_{k,\alpha}' \rho' {\varsigma_{k,\beta}'}^\dagger V^\dagger {\varsigma_{d,\nu}'}^\dagger \right).
\end{equation}\\
To express the quartic polynomial in a form that can be reconstructed from circuit evaluations, we group similar monomials according to repeated and mixed indices and define the following coefficients

\begin{align}
    &\rr \left(\varsigma_{d,\mu}', \varsigma_{k,\alpha}' \right) \coloneqq \A_{\mu \mu \alpha \alpha}, \label{R_coefficient_1} \\[0.2cm]
    &\rr\left( \varsigma_{d,\mu}', \varsigma_{k,\alpha}', \varsigma_{k,\beta}' \right) \coloneqq \A_{\mu \mu \alpha \beta} + \A_{\mu \mu \beta \alpha}, \quad \alpha \neq \beta, \label{R_coefficient_2} \\[0.2cm]
    &\rr\left( \varsigma_{d,\mu}', \varsigma_{d,\nu}', \varsigma_{k,\alpha}' \right) \coloneqq \A_{\mu \nu \alpha \alpha} + \A_{\nu \mu \alpha \alpha}, \quad \mu \neq \nu, \label{R_coefficient_3} \\[0.2cm]
    &\rr\left( \varsigma_{d,\mu}', \varsigma_{d,\nu}', \varsigma_{k,\alpha}', \varsigma_{k,\beta}' \right)  \coloneqq \A_{\mu \nu \alpha \beta} +  \A_{\mu \nu \beta \alpha} + \A_{\nu \mu \alpha \beta} \nonumber \\ 
    & \hspace{3.5cm} + \A_{\nu \mu \beta \alpha}, \quad  \mu \neq \nu, \ \alpha \neq \beta. \label{R_coefficient_4}
\end{align}\\
Using these coefficients, the two-gate cost function becomes 

\begin{align}\label{tgfqs_cost_function}
\begin{split}
    \expval{M}_{\q_d, \q_k} &= \sum_{\mu, \alpha=0}^3 q_{d, \mu}^2 q_{k, \alpha}^2 \rr\left( \varsigma_{d,\mu}', \varsigma_{k,\alpha} ' \right) \\
    & + \sum_{\substack{\mu, \nu, \alpha=0  \\ \mu \neq \nu}}^3  q_{d, \mu}  q_{d, \nu}  q_{k, \alpha}^2 \rr\left( \varsigma_{d,\mu}', \varsigma_{d,\nu}', \varsigma_{k,\alpha}' \right) \\
    & + \sum_{\substack{\mu, \alpha, \beta =0 \\ \alpha \neq \beta}}^3   q_{d, \mu}^2  q_{k, \alpha}  q_{k, \beta} \rr\left( \varsigma_{d,\mu}', \varsigma_{k,\alpha}', \varsigma_{k,\beta}' \right) \\
    & +  \sum_{\substack{\mu, \nu, \alpha, \beta  =0\\ \mu \neq \nu \\ \alpha \neq \beta}}^3   q_{d, \mu} q_{d, \nu}   q_{k, \alpha}  q_{k, \beta} \rr\left( \varsigma_{d,\mu}', \varsigma_{d,\nu}', \varsigma_{k,\alpha}', \varsigma_{k,\beta}' \right).
\end{split}
\end{align}

Unlike the single-gate Fraxis and FQS updates, which yield quadratic local cost functions, the two-gate local cost function is quartic. To evaluate the coefficients~(\ref{R_coefficient_1})--(\ref{R_coefficient_4}) in terms of expectation values, similarly to single-gate Fraxis and FQS, we first define 

\begin{equation}
    \trt(O_d, O_k) \coloneqq \Tr(M' O_d V O_k \rho' O_k^\dagger V^\dagger O_d ^\dagger),
\end{equation}\\
where $O_d$ and $O_k$ denote the operators that are inserted in place of the $d$-th and $k$-th parameterized gates, respectively. The coefficients $\rr\bigl( \varsigma_{d,\mu}', \varsigma_{k,\alpha}' \bigr)$ in Eq.~(\ref{R_coefficient_1}) can be computed as follows

\begin{equation}
    \rr\left( \varsigma_{d,\mu}', \varsigma_{k,\alpha}' \right) = \trt\left( \varsigma_{d,\mu}', \varsigma_{k,\alpha}' \right).
\end{equation}\\
To compute the mixed terms coefficients in Eqs.~(\ref{R_coefficient_2})--(\ref{R_coefficient_4}) we first define 

\begin{equation}
    \varsigma_{(\mu + \nu)}' \coloneqq \frac{\varsigma_{\mu}' + \varsigma_{\nu}'}{\sqrt{2}}, \quad \mu \neq \nu.
\end{equation}
This is similar to the computation of the off-diagonal elements in single-gate Fraxis and FQS~\cite{fraxis, fqs}. The cubic coefficients are then given by

\begin{widetext}
\begin{align}\label{R2_term_equation}
    \rr\left( \varsigma_{d, \mu}', \varsigma_{k, \alpha}', \varsigma_{k, \beta}' \right) &= 2\cdot \trt\left( \varsigma_{d, \mu}', \varsigma_{k, (\alpha + \beta)}' \right) -  \trt\left( \varsigma_{d, \mu}', \varsigma_{k, \alpha}' \right) - \trt\left( \varsigma_{d, \mu}', \varsigma_{k, \beta}' \right), \quad \alpha \neq \beta, \\[0.2cm]
    \rr\left( \varsigma_{d, \mu}', \varsigma_{d, \nu}', \varsigma_{k, \alpha}' \right) &= 2 \cdot \trt\left( \varsigma_{d, (\mu + \nu)}', \varsigma_{k, \alpha}' \right) -  \trt\left( \varsigma_{d, \mu}', \varsigma_{k, \alpha}' \right) - \trt\left( \varsigma_{d, \nu}', \varsigma_{k, \alpha}' \right), \quad \mu \neq \nu.
\end{align}
\end{widetext}
Finally, the quartic coefficients $\rr\left( \varsigma_{d,\mu}', \varsigma_{d,\nu}', \varsigma_{k,\alpha}', \varsigma_{k,\beta}' \right)$ can be computed as follows

\begin{widetext}
\begin{align}\label{R4_term_equation}
\begin{split}
&\rr\left( \varsigma_{d,\mu}', \varsigma_{d,\nu}', \varsigma_{k,\alpha}', \varsigma_{k,\beta}' \right) = 4 \cdot \trt\left( \varsigma_{d, (\mu+\nu)}', \varsigma_{k, (\alpha + \beta)}' \right)  +  \trt\left( \varsigma_{d, \mu}', \varsigma_{k, \alpha}' \right) +  \trt\left( \varsigma_{d, \mu}', \varsigma_{k, \beta}' \right) +  \trt\left( \varsigma_{d, \nu}', \varsigma_{k, \alpha}' \right) +  \trt\left( \varsigma_{d, \nu}', \varsigma_{k, \beta}' \right) \\[0.2cm]
&\quad \quad \quad - 2\cdot \trt\left( \varsigma_{d, \mu}', \varsigma_{k, (\alpha + \beta)}' \right) - 2\cdot \trt\left( \varsigma_{d, \nu}', \varsigma_{k, (\alpha + \beta)}' \right) - 2\cdot \trt\left( \varsigma_{d, (\mu + \nu)}', \varsigma_{k, \alpha}' \right) -  2\cdot \trt\left( \varsigma_{d, (\mu + \nu)}', \varsigma_{k, \beta}' \right), \quad \mu \neq \nu, \ \alpha \neq \beta.
\end{split}
\end{align}
\end{widetext}

\begin{algorithm*}
\vspace{0.1cm}
\caption{Two-Gate FQS (TGFQS)}\label{tgfqs_algorithm}
\begin{algorithmic}[1]
\State \textbf{Inputs}: A Parameterized Quantum Circuit $U$ with fixed architecture, Hermitian measurement operator $M$ as the cost function, stopping criterion, a classical optimizer (e.g., SLSQP), and a gate pairing strategy (e.g., random or linear).
\State Initialize the parameters, $\q_i \in \mathbb{R}^4$ with $\abs{\q_i}=1$ for $i = 1, \ldots, D$.
\Repeat
    \State Construct a gate pair sequence $\mathcal{P} = \{(d_1, k_1),(d_2, k_2),\ldots,(d_{D/2}, k_{D/2})\}$, according to the selected pairing strategy, where each index appears exactly once.
    \For{Gate pair $(d,k) \in \mathcal{P}$}
        \State Fix all parameterized gates except the $d$-th and $k$-th ones.
        \State Compute the coefficients $\rr$ appearing in Eq.~(\ref{tgfqs_cost_function}).
        \State Find optimal $\q_d^*=(q_{d,0}^*, q_{d,1}^*, q_{d,2}^*, q_{d,3}^*)^T$ and $\q_k^* =(q_{k,0}^*, q_{k,1}^*, q_{k,2}^*, q_{k,3}^*)^T$ that minimize the cost function $\expval{M}_{\q_d,\q_k}$ with the constraints $\abs{\q_d} = \abs{\q_k} = 1$ and using a classical optimizer.
        \If{$\expval{M}_{\q_d^*, \q_k^*} < \expval{M}_{prev}$}
        \State $\q_d \leftarrow \q_d^*$
        \State $\q_k \leftarrow \q_k^*$
        \EndIf
    \EndFor
\Until{stopping criterion is met.}
\end{algorithmic}\vspace{0.1cm}
\end{algorithm*}

In Appendix~\ref{R_2_and_R_4_trace_calculation_appendix}, we provide the derivations for Eqs.~(\ref{R2_term_equation})--(\ref{R4_term_equation}). The local cost function for the TGF is obtained by setting the first component of the unit quaternions to zero 

\begin{equation}
    \q_d = (0, \n_d), \quad \q_k = (0, \n_k),
\end{equation}\\
where $\n_d, \n_k \in \mathbb{R}^3$ and $\abs{\n_d} = \abs{\n_k} = 1$. Consequently, only the indices $\{1,2,3\}$ contribute to constructing the local two-gate cost function, reducing the number of required components and expectation values. Similarly to single-gate Fraxis and FQS, where Fraxis is a special case of FQS, the TGF is a special case of TGFQS. 

The local two-gate cost function $\expval{M}_{\q_d, \q_k}$ is then minimized by solving a constrained classical optimization problem

\begin{align}
\begin{split}
     (\q_d^*, \q_k^*) &= \underset{\q_d, \q_k}{\arg \min} \expval{M}_{\q_d,\q_k}, \\[0.1cm]
     & \ s.t. \ \abs{\q_d} = \abs{\q_k} = 1,
\end{split}
\end{align}\\
where $\q_d^*$ and $\q_k^*$ represent the optimal quaternions. The TGF solves the same problem but with reduced parameterization via $\q_d = (0, \n_d)$ and $\q_k = (0, \n_k)$. Since the two-gate cost function $\expval{M}_{\q_d,\q_k}$ is quartic, an analytic closed-form solution is generally not available. We therefore solve for the optimal quaternions  $\q_d^*, \q_k^*$ numerically using the SLSQP optimizer~\cite{SLSQP_ref1, SLSQP_ref2}. Other classical constrained optimizers, such as COBYLA~\cite{cobyla_optimizer}, could also be used. 

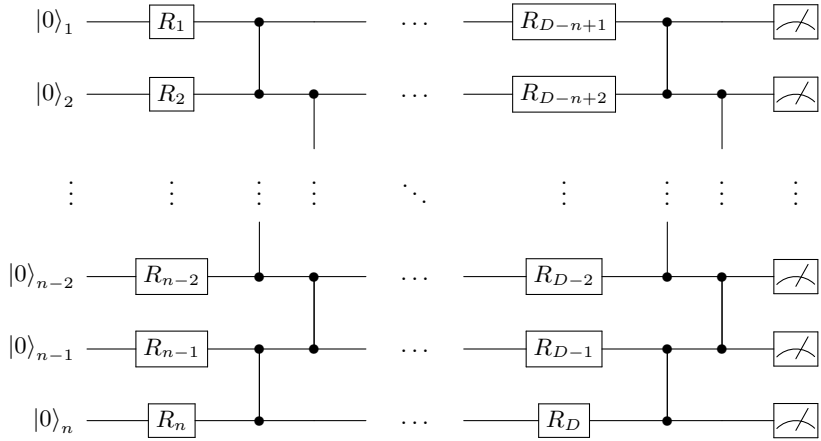
\begin{figure*}
\vspace{1cm}
\centering
\[
\Qcircuit @C=2em @R=1.5em {
\lstick{\ket{0}_1} & \gate{R_1} & \ctrl{1} & \qw & \qw & \ldots & & \gate{R_{D-n+1}} & \ctrl{1} & \qw &  \meter \\
\lstick{\ket{0}_2} & \gate{R_2} & \ctrl{-1}    & \ctrl{1} & \qw &  \ldots & & \gate{R_{D-n + 2}} & \ctrl{-1} &  \ctrl{1} & \meter \\
\lstick{}    &          &  &   & & & & & &    \\
\lstick{\vdots}    & \vdots          &  \vdots  & \vdots   &  & \ddots &  & \vdots &  \vdots & \vdots & \vdots \\
\lstick{}    &     & & & & &     &  &   &  \\
\lstick{\ket{0}_{n-2}} & \gate{R_{n-2}} & \ctrl{-1} & \ctrl{1} & \qw &\ldots & & \gate{R_{D-2}}  &\ctrl{-1} & \ctrl{1} & \meter \\
\lstick{\ket{0}_{n-1}} & \gate{R_{n-1}} & \ctrl{1} & \ctrl{-1} & \qw & \ldots & &  \gate{R_{D-1}} & \ctrl{1} & \ctrl{-1} & \meter \\
\lstick{\ket{0}_n} & \gate{R_n} & \ctrl{-1} & \qw & \qw & \ldots & &\gate{R_D} & \ctrl{-1} & \qw & \meter
}
\]
\caption{Hardware-efficient ansatz consisting of parameterized single-qubit gates $R_i$ for $ i =1,2,\ldots, D$ followed by an entangling layer consisting of controlled-Z gates for every qubit pair, repeated over $L$ layers for $n$ qubits.}
\label{ansatz_circuit}
\end{figure*}

Let $D$ denote the total number of parameterized single-qubit gates in the PQC. We assume that a hardware-efficient ansatz depicted in Fig.~\ref{ansatz_circuit} is utilized in optimization, where parameterized single-qubit gates act on every qubit and are followed by an entangling layer consisting of fixed two-qubit gates, such as CNOTs or controlled-Z gates. This structure is then repeated over $L$ times, and the total number of parameterized gates is $D = Ln$. During a single optimization iteration, the set of parameterized gate indices $\{1,2,\ldots, D\}$ can be partitioned into disjoint pairs according to a chosen pairing strategy. That is, the set of gate indices $\{1,2,\ldots,D\}$ is partitioned into $D/2$ disjoint pairs

\begin{equation}
 \mathcal{P} = \{(d_1, k_1),(d_2, k_2),\ldots,(d_{D/2}, k_{D/2})\}    
\end{equation}\\ 
by the given gate pairing strategy. Here, the iteration refers to updating all parameterized gates in the PQC once according to a selected pair sequence. As the gates are partitioned into a total of $D/2$ pairs, we assume that $D=Ln$ is even, so that all parameterized gates can be partitioned into pairs without any unpaired gates. This holds when either $n$ or $L$ is even. The gate pairs can be chosen randomly or according to simple deterministic heuristics. In this work, we consider the following gate pairing strategies: linear, random, opposite, and half-shifted gate pairings. These are defined as follows

\begin{align}
    \mathcal{P}_{\mathrm{linear}} &= \{(1,2), (3,4), \ldots, (D-1, D)\}, \\[0.2cm]
    \mathcal{P}_{\mathrm{random}} &= \{(\pi(1), \pi(2)), \ldots, (\pi(D-1), \pi(D))\}, \\[0.2cm]
    \mathcal{P}_{\mathrm{opp}} &= \{(1, D), (2, D-1), \ldots, (D/2, D/2+1)\}, \\[0.2cm]
    \mathcal{P}_{\mathrm{hs}} &= \{(1, D/2+1), (2, D/2+2),\ldots,(D/2, D)\}.
\end{align}
Here, $\pi(i)$ denotes the $i$-th element of a random permutation of the index set $\{1,2,\ldots,D\}$. For example, when $D=10$, we have gate pair sequences as follows

\begin{align}
    \mathcal{P}_{\mathrm{linear}} &= \{(1,2), (3,4), (5,6), (7,8), (9,10)\}, \\[0.2cm]
    \mathcal{P}_{\mathrm{random}} &= \{(1,4), (7,2), (10,8), (3,6), (5,9)\}, \\[0.2cm]
    \mathcal{P}_{\mathrm{opp}} &= \{(1,10), (2,9), (3,8), (4,7), (5,6)\}, \\[0.2cm]
    \mathcal{P}_{\mathrm{hs}} &= \{(1,6), (2,7), (3,8), (4,9), (5,10)\}.
\end{align}\\
For the random gate pairing strategy, a new random permutation is generated at the beginning of each iteration.

After optimizing the gate pair, if the optimal quaternions $\q_d^*$ and $\q_k^*$ lower the cost function relative to the previous value, the parameters of the $d$-th and $k$-th parameterized gates are updated. Otherwise, they stay the same. The TGFQS is described in Algorithm~\ref{tgfqs_algorithm}. To compare the cost of each optimizer, Table~\ref{algo_table} summarizes the number of circuit evaluations required to construct the local cost function and perform the corresponding classical optimization for a single parameterized single-qubit gate update. TGF and TGFQS require a total of 36 and 100 circuit evaluations to update one gate pair, respectively. This translates to 18 and 50 circuit evaluations on average per updated gate.

\begin{table}
\centering
\scalebox{1.1}{%
\begin{tabular}{|c |c | c | c | c |} 
 \hline
 Algorithm & Fraxis & FQS & TGF & TGFQS  \\ 
 \hline
 Circuit evaluations & 6 & 10 & 18 & 50 \\
 \hline
\end{tabular}}
\caption{Average number of circuit evaluations required per updated gate to construct the local cost function for each optimizer. For TGF and TGFQS, one gate pair update requires 36 and 100 circuit evaluations, respectively, corresponding to 18 and 50 circuit evaluations per updated gate.} 
\label{algo_table}
\end{table}

\section{Results}\label{results_section}

In this section, we present a set of numerical experiments with the proposed TGF and TGFQS methods and compare the performance to their standard versions, Fraxis and FQS, respectively. First, we go through the results of the spin Hamiltonians. We used a 4-qubit Fermi-Hubbard Hamiltonian on a $1\times2$ lattice and a transverse-field Ising model with 8, 10, and 12 qubits. Then, we show the results for the 12-qubit lithium hydride (LiH) and 14-qubit beryllium hydride (BeH$_2$) molecular Hamiltonians. Finally, we evaluate the ability of TGF and TGFQS to prepare target quantum states by maximizing fidelity to a randomly sampled quantum state using 6 qubits.

In all simulations, we used the PennyLane Python package~\cite{pennylane}. For the ansatz circuit, we employed a hardware-efficient ansatz depicted in Fig.~\ref{ansatz_circuit}. For standard Fraxis and FQS, the gate update sequence proceeds from the top left gate to the bottom right gate of the circuit, layer by layer, as depicted in Fig.~\ref{ansatz_circuit}. Once all parameterized single-qubit gates in the circuit have been updated, we begin a new iteration and continue the optimization until a stopping criterion is met. In this work, we use a fixed number of iterations as a stopping criterion. One iteration is defined as updating all parameterized single-qubit gates in the PQC once according to the given sequence. At the beginning of each run, we initialized the parameters for Fraxis, FQS, TGF, and TGFQS using uniform spherical distributions in three and four dimensions to sample unit axes and unit quaternions, respectively. For the two-gate optimizers, the constrained local subproblem was solved by using SLSQP. We tested several different gate pairing strategies, namely, linear, random, opposite, and half-shifted gate pairings. Unless otherwise stated, we used a noiseless statevector simulator in the experiments. 

We assume that the total number of parameterized gates is even, i.e., $ D=Ln$ is even. This ensures that all parameterized gates can be partitioned into disjoint pairs without any unpaired gates. In this work, we use an even number of qubits for all experiments. To facilitate visual comparison of single- and two-gate methods, standard Fraxis and FQS are shown in red, whereas two-gate pairing strategies are shown in blue (linear), green (random), black (opposite), and orange (half-shifted).

\begin{figure}
    \centering
    \includegraphics[width=0.95\linewidth]{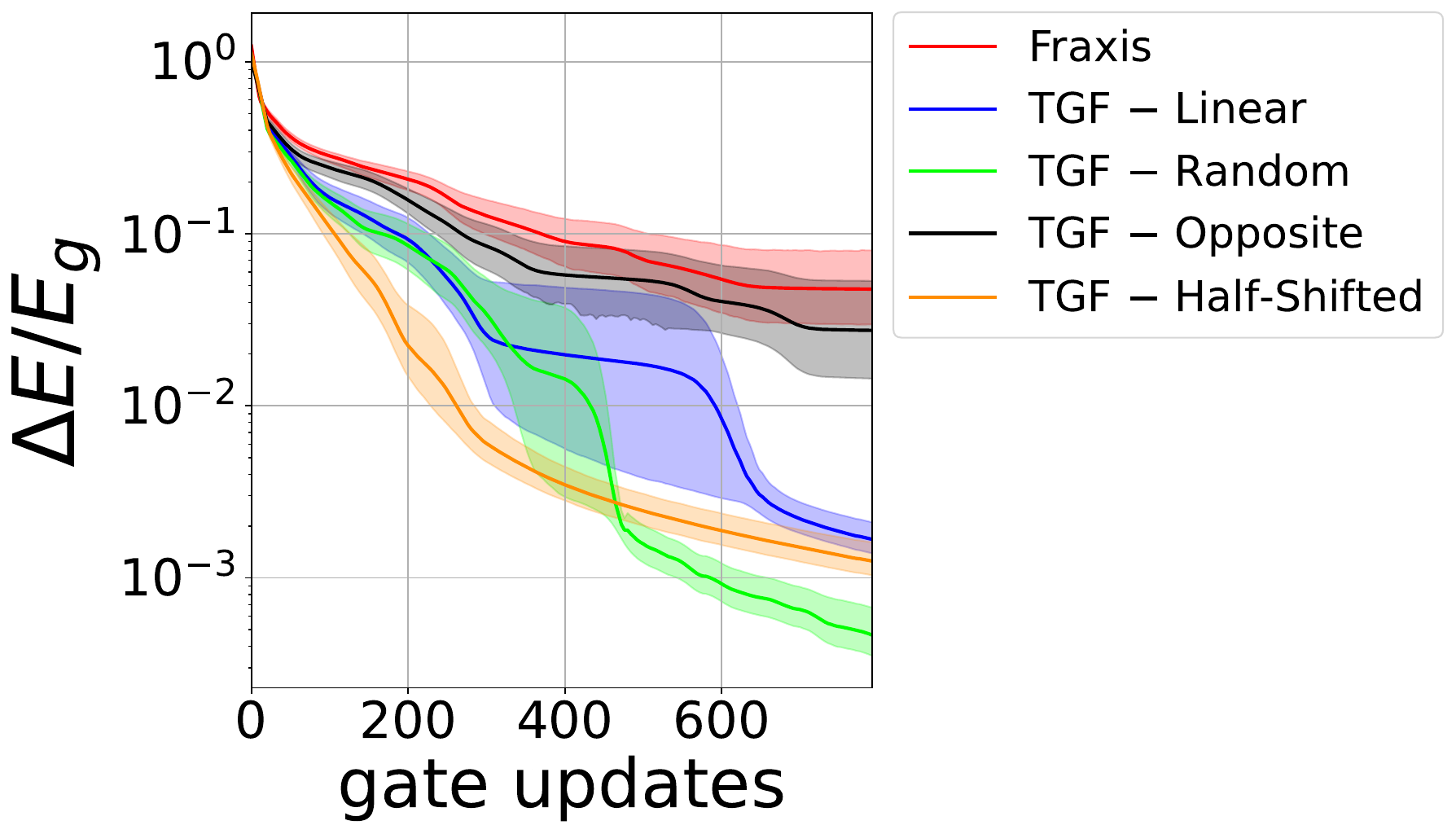}\\[0.5cm]
    \includegraphics[width=0.99\linewidth]{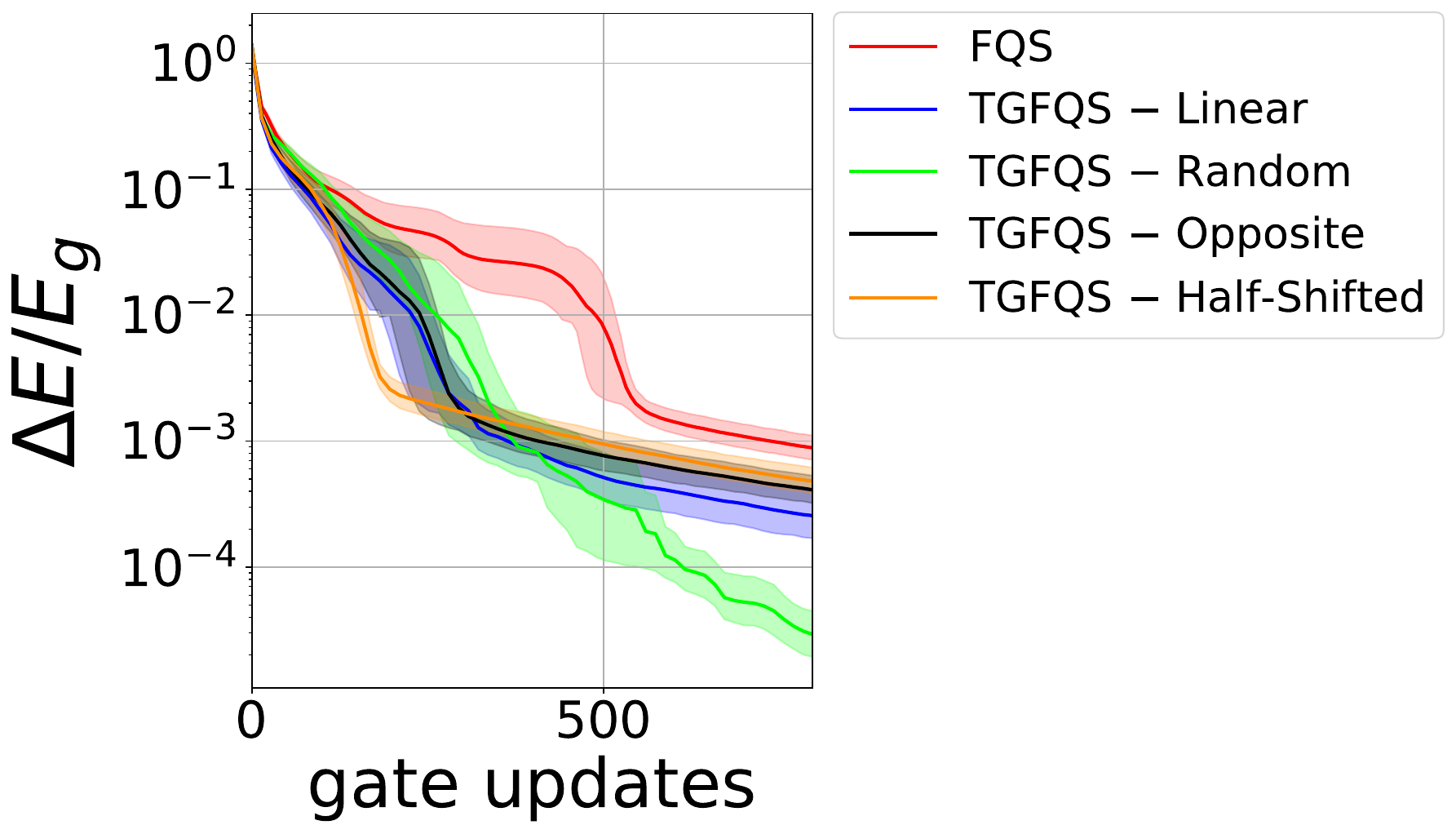}
    \cprotect\caption{Results for 4-qubit Fermi-Hubbard model on a $1\times2$ lattice for standard Fraxis and FQS (red) as well as TGF and TGFQS with the following gate pairs: linear (blue), random (green), opposite (black), and half-shifted (orange). In both figures, the number of layers was set to $L=4$, each line represents a mean of 20 runs, and the shaded areas are 68\% confidence intervals around the mean. The figures are shown on a semi-log scale, with the vertical axis showing the relative error with respect to the ground state energy.}
    \label{fermi_hubbard_results}
\end{figure}

\begin{figure}
    \centering
    \includegraphics[width=0.8\linewidth]{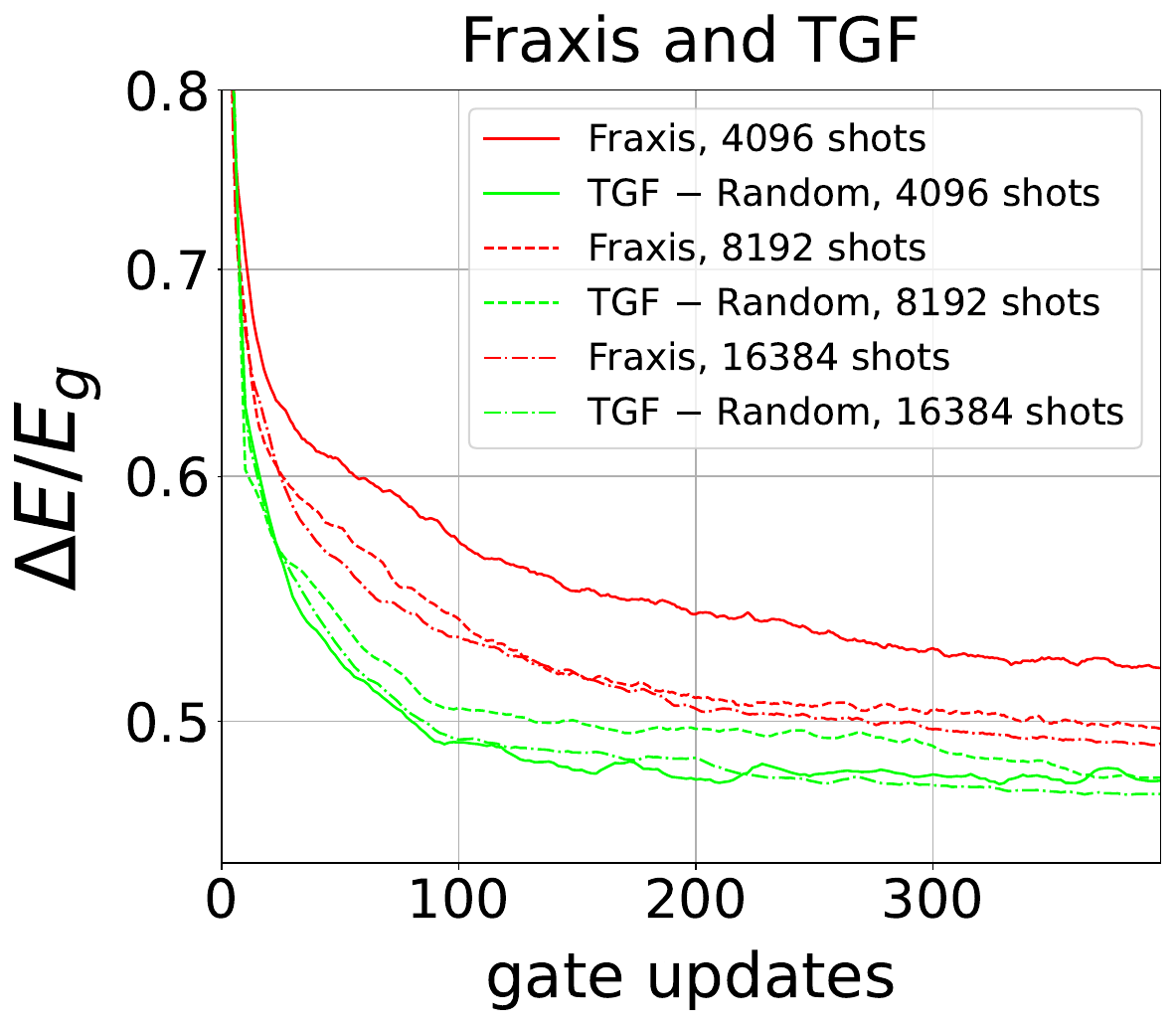}\\[0.5cm]
    \includegraphics[width=0.8\linewidth]{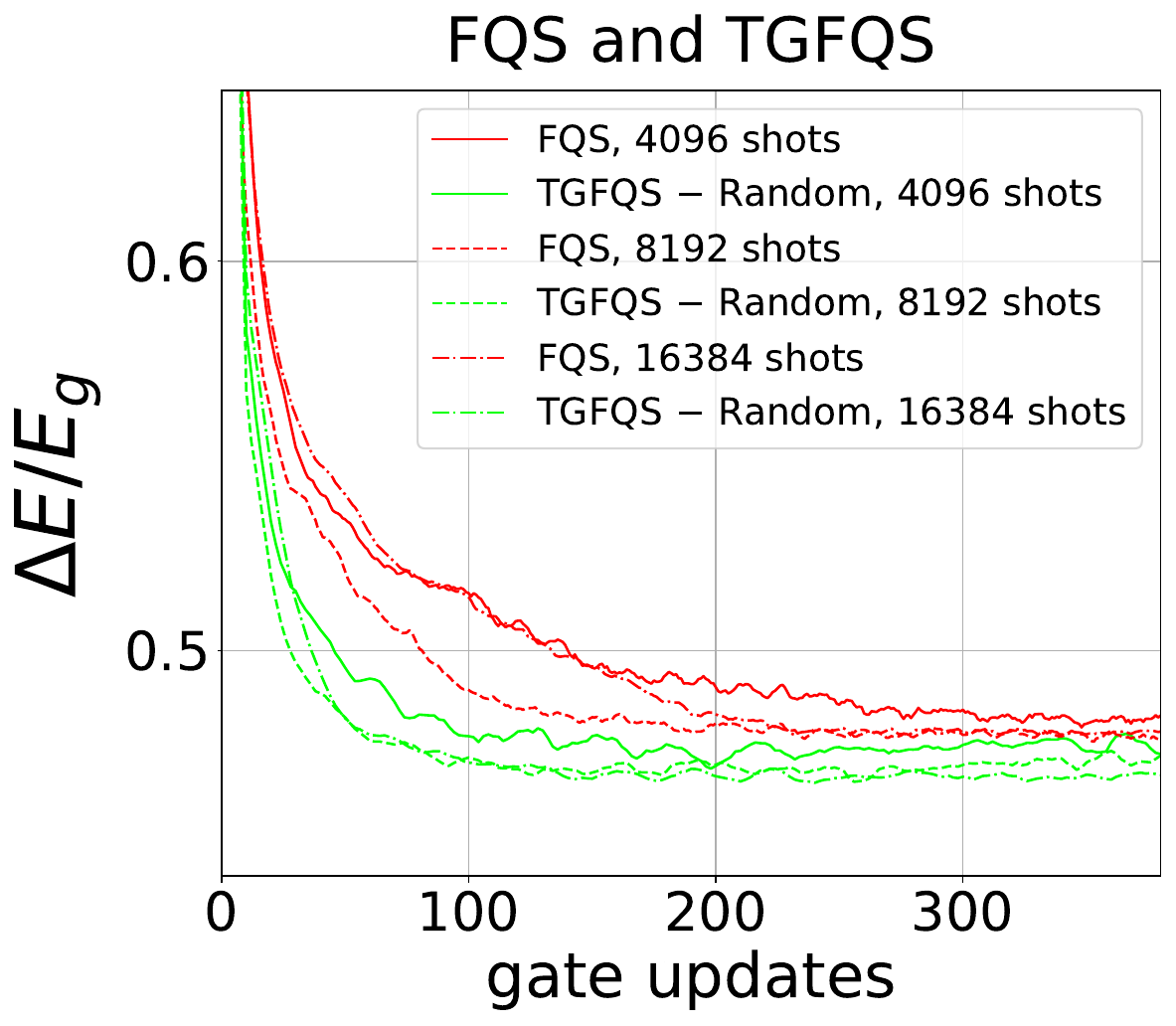}
    \cprotect\caption{Results for 4-qubit Fermi-Hubbard model on a $1\times2$ lattice for standard Fraxis and FQS (red) as well as TGF and TGFQS with random gate pairing (green). In both figures, the number of layers was set to $L=2$, and each line represents a mean of 20 runs. The line styles indicate the number of shots: solid for 4096 shots, dashed for 8192 shots, and dash-dotted for 16384 shots. The vertical axis shows the relative error with respect to the ground state energy.}
    \label{fermi_hubbard_results_noisy}
\end{figure}

\subsection{Spin Hamiltonians}

In this section, we examine the performance of Fraxis, FQS, TGF, and TGFQS with different spin Hamiltonians. We present results for the 4-qubit Fermi-Hubbard model~\cite{fermi_hubbard_model} and the transverse-field Ising model (TFIM)~\cite{TFIM} with 8, 10, and 12 qubits, while keeping the number of layers fixed to $L=4$.

\begin{figure*}
    \centering
    \includegraphics[width=0.95\linewidth]{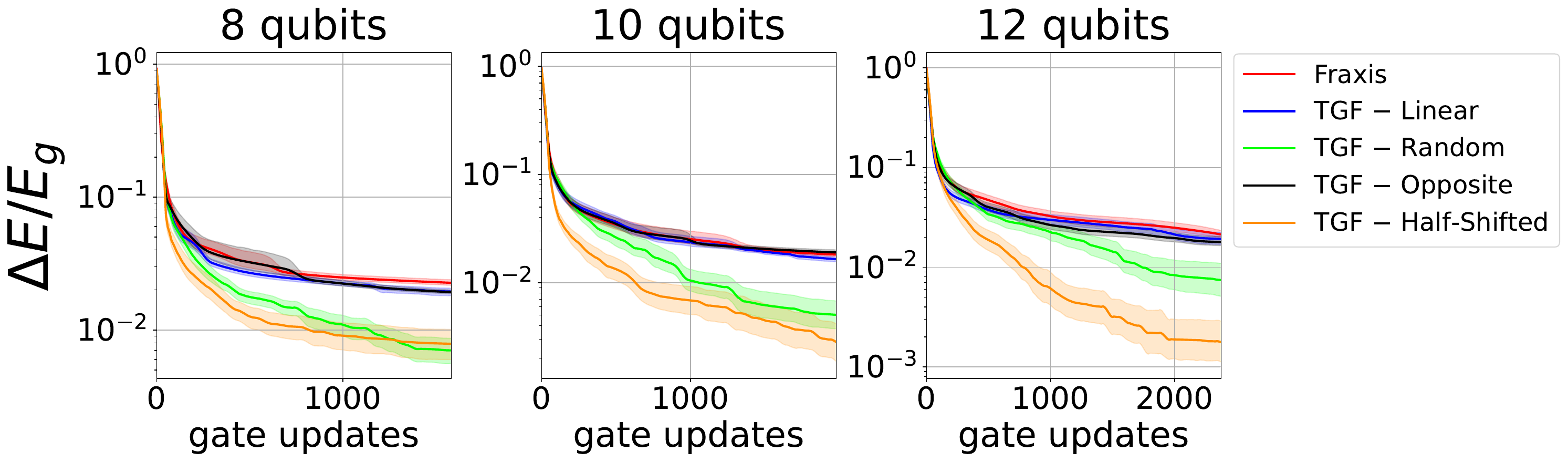}
    \includegraphics[width=0.99\linewidth]{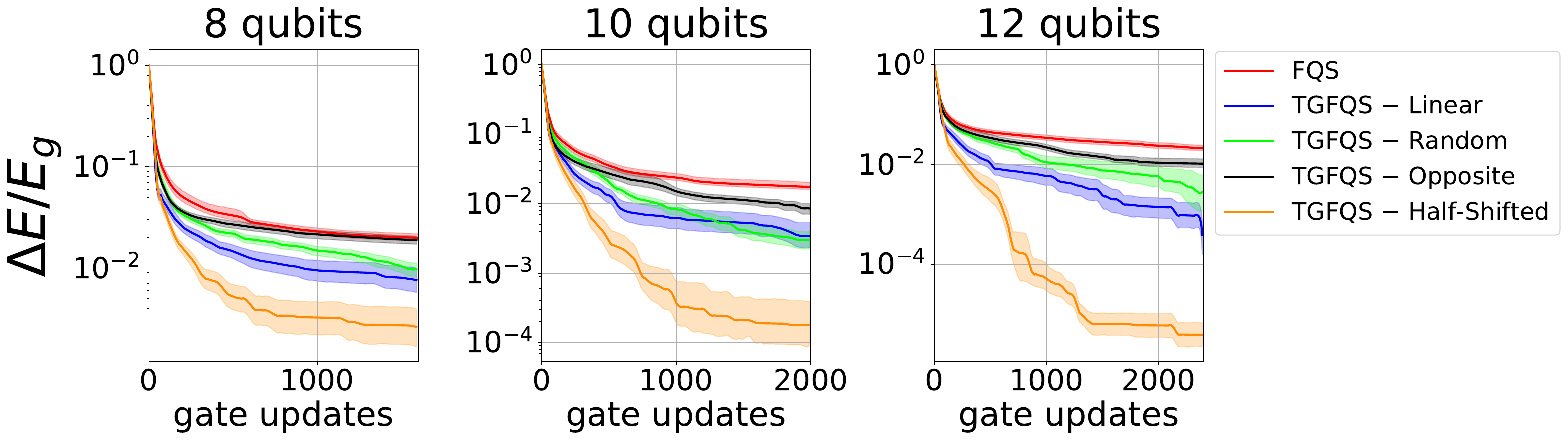}
    \cprotect\caption{Results for 8-, 10-, and 12-qubit TFIM Hamiltonian as a function of gate updates performed while setting the number of layers to $L=4$ using single-gate optimizers Fraxis and FQS (red) and two-gate optimizers TGF and TGFQS with linear (blue), random (green), opposite (black), and half-shifted (orange) gate pairings. In all figures, each line represents a mean of 20 runs, and the shaded areas are 68\% confidence intervals around the mean. The figures are shown on a semi-log scale, with the vertical axis showing the relative error with respect to the ground state energy.}
    \label{TFIM_results}
\end{figure*}

\begin{figure}
    \centering
    \includegraphics[width=0.85\linewidth]{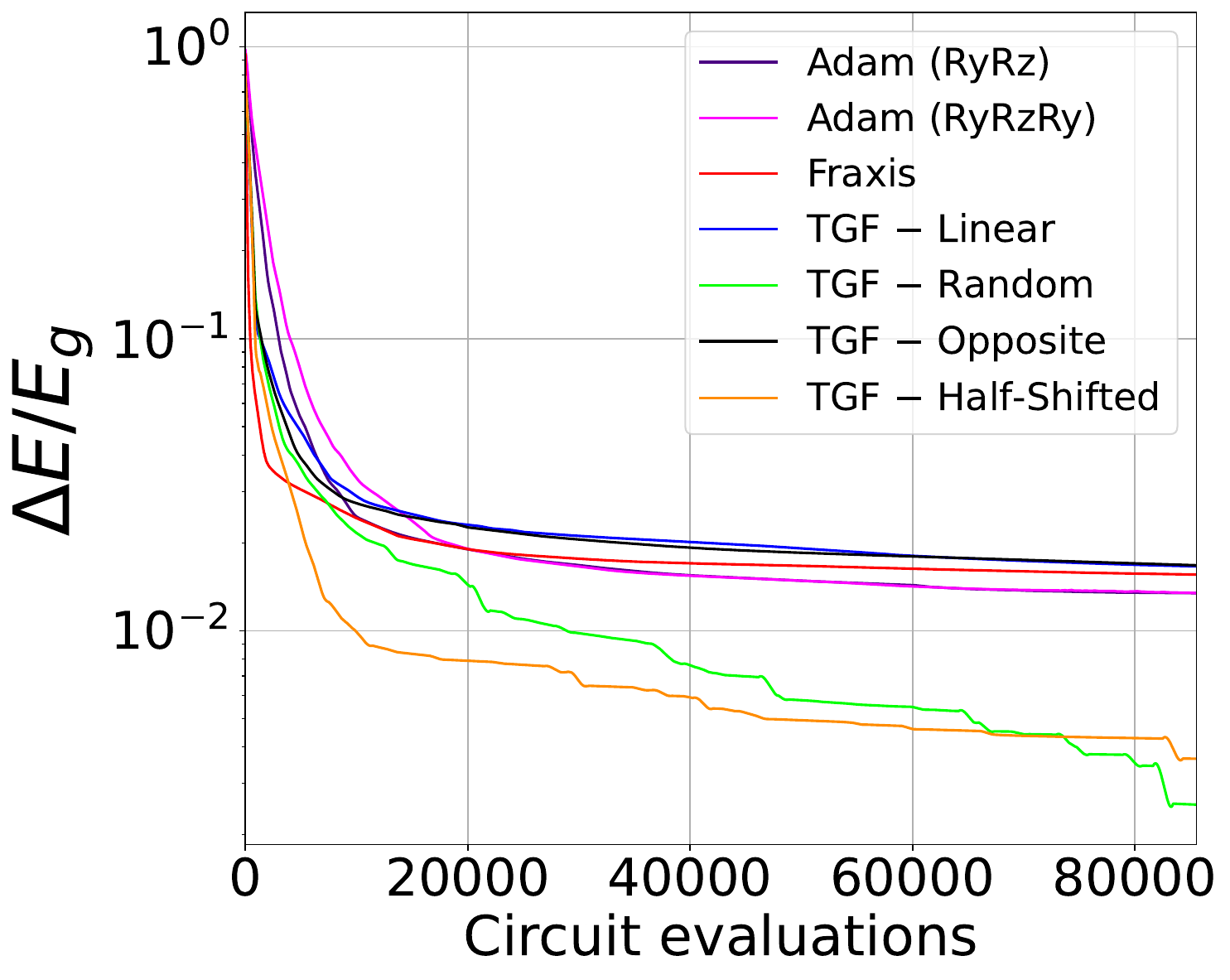}
    \\[0.2cm]
    \includegraphics[width=0.85\linewidth]{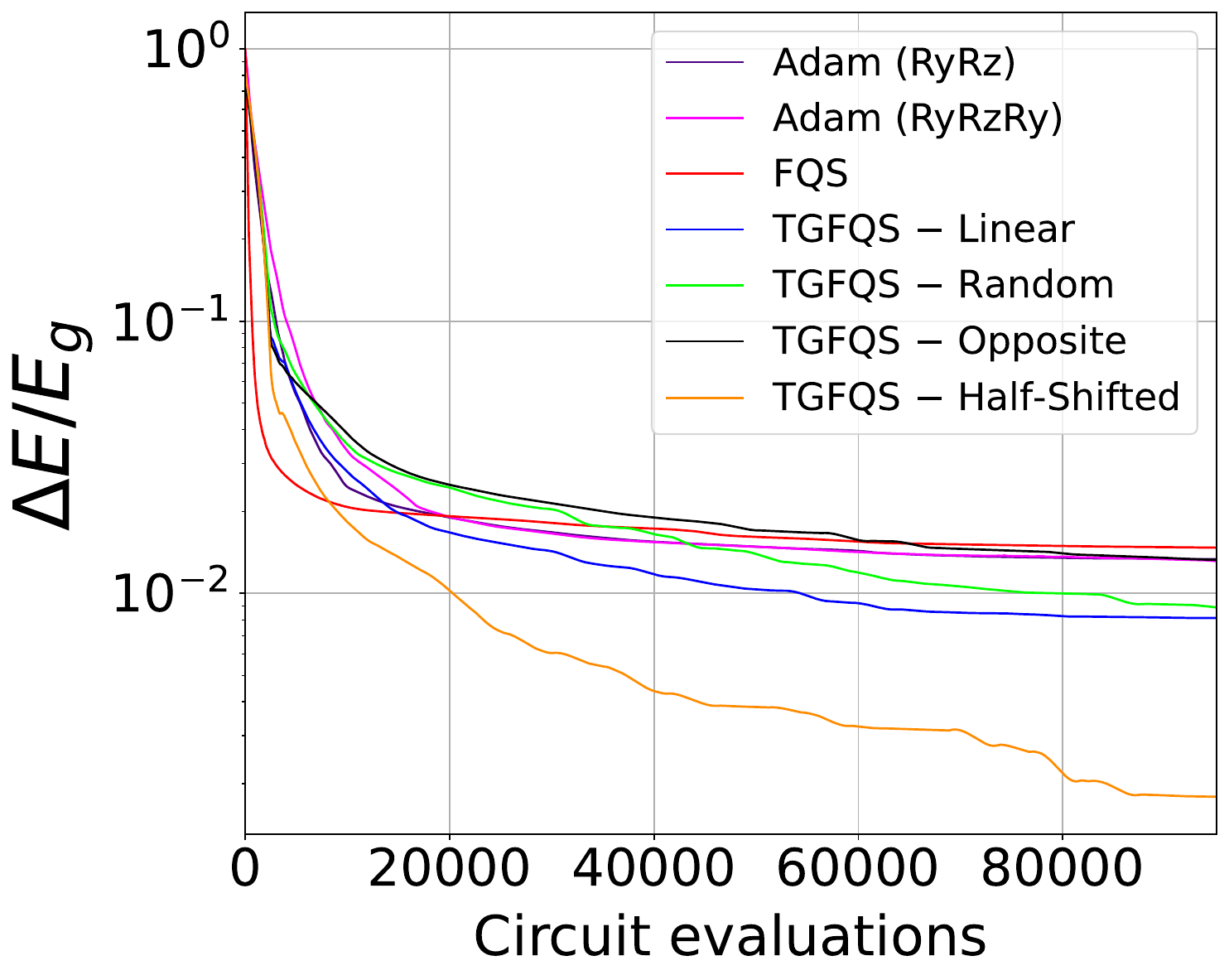}
    \cprotect\caption{Results for 8-qubit TFIM Hamiltonian as a function of circuit evaluations performed while setting the number of layers to $L=4$. Gradient-based method, Adam with $R_Y R_Z$ and $R_Y R_Z R_Y$ gate decompositions were employed in addition to the single-gate optimizers Fraxis and FQS (red), and two-gate optimizers TGF and TGFQS with linear (blue), random (green), opposite (black), and half-shifted (orange) gate pairings. The learning rate for the Adam optimizer was set to 0.05. In both figures, each line represents the mean of 20 runs.}
    \label{TFIM_results_circuit_evals_8Q_4L}
\end{figure}

\begin{figure}
    \centering
    \includegraphics[width=0.85\linewidth]{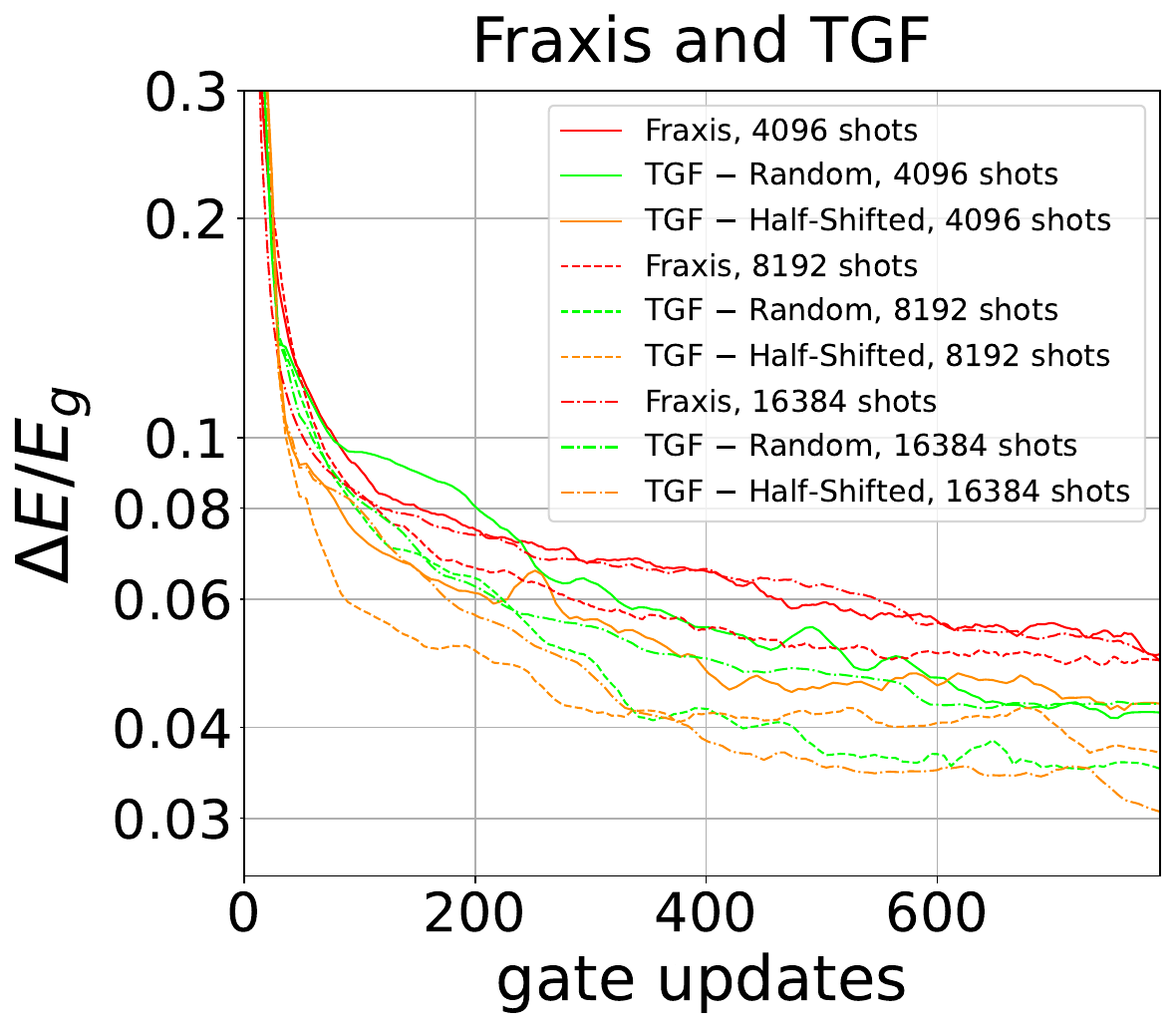}
    \centering
    \includegraphics[width=0.9\linewidth]{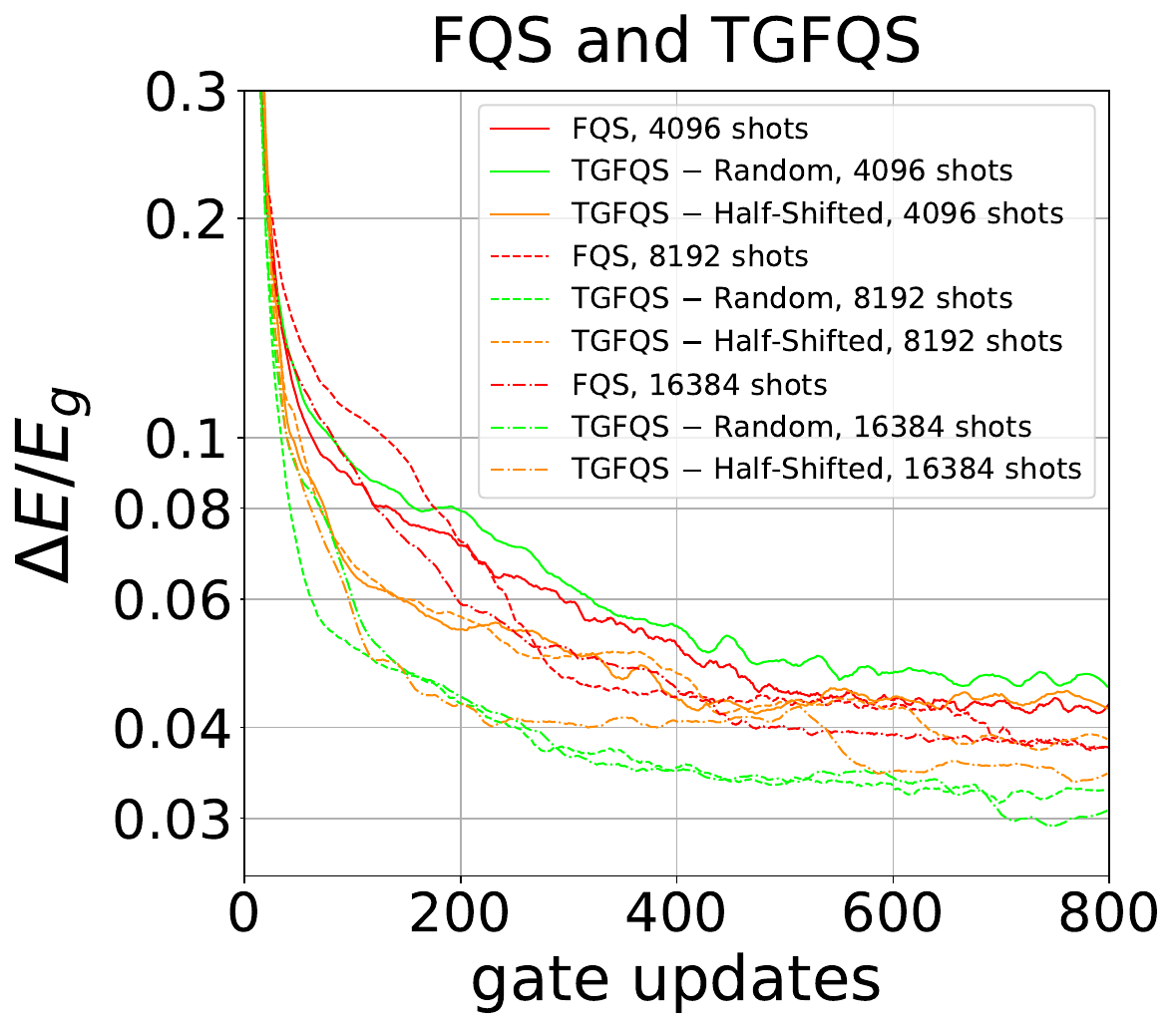}
    \cprotect\caption{Results for 8-qubit TFIM Hamiltonian as a function of gate updates performed while setting the number of layers to $L=2$ using single-gate optimizers Fraxis and FQS (red) and two-gate optimizers TGF and TGFQS with random (green) and half-shifted (orange) gate pairings. In all figures, each line represents a mean of 20 runs, and the line styles indicate the number of shots: solid for 4096 shots, dashed for 8192 shots, and dash-dotted for 16384 shots. The vertical axis shows the relative error with respect to the ground state energy.}
    \label{TFIM_results_8Q_noisy_2L_noisy}
\end{figure}

\subsubsection{Fermi--Hubbard Model}

For the first spin Hamiltonian benchmark, we consider the one-dimensional Fermi--Hubbard (FH) model in a $ 1\times 2$ lattice. The FH model describes how the fermions interact on a lattice and has been used in quantum simulations and computing~\cite{FH_quantum_simulation, FH_opt_lattice, FH_nisq_quantum_sim, FH_quantu_sim_silicon, FH_quantum_computing}. The FH Hamiltonian is given as follows
\begin{equation}
    H_{FH} = -t \sum_{<i,j>, \sigma} a_{i\sigma}^\dagger a_{j\sigma} + U_{C} \sum_i n_{i\uparrow} n_{i \downarrow}.
\end{equation}
Here, $t$ and $U_{C}$ denote the kinetic term, which is usually called the hopping term, and the on-site Coulomb potential between the fermions, respectively. The operators $a_{i\sigma}^\dagger$ and $a_{j\sigma}$ denote fermionic creation and annihilation operators, and $<i,j>$ are the indices of the neighboring lattice sites. To form the cost function from the FH Hamiltonian to the Pauli basis, we apply Jordan--Wigner mapping~\cite{jordan_wigner} to the fermionic operators. We extract the corresponding Hamiltonian by using the PennyLane Python package~\cite{pennylane} with given parameters for $t$ and $U_{C}$. In this work, we set $t = U_{C} =0.75$.

The results for the 4-qubit FH model after 50 iterations and 20 runs for Fraxis, FQS, and each gate pairing strategy of TGF and TGFQS are shown in Fig.~\ref{fermi_hubbard_results}. TGF and TGFQS achieve a lower relative error at the end of optimization than Fraxis and FQS, respectively, across all gate pairing strategies. Among the tested gate pairing strategies, the random gate pairing yields the lowest relative error at the end of optimization. Linear, half-shifted, and opposite gate pairings also obtain a lower relative error of the mean than their single-gate counterparts. Depending on the pairing strategy, the improvement ranges from modest gains to approximately two orders of magnitude lower relative error at the end of optimization.

We additionally examined the performance of Fraxis, FQS, and two-gate optimizers with a random gate pairing strategy in the presence of finite measurement accuracy. The finite measurement accuracy was modeled by estimating each Hamiltonian term with shot noise using 4096, 8192, and 16384 shots and setting the number of layers to $L=2$. The results are shown in Fig.~\ref{fermi_hubbard_results_noisy}. For all tested shot counts, both TGF and TGFQS achieve a lower relative error across 20 runs compared to Fraxis and FQS, respectively. The performance gap is smaller than in the noiseless statevector simulations. We also observed that this advantage is obtained with the shallow circuit depth, and increasing the number of layers does not improve the performance of TGF and TGFQS relative to Fraxis and FQS, respectively.

\subsubsection{Transverse--Field Ising Model}

For the second spin Hamiltonian benchmark, we chose the one-dimensional transverse-field Ising model~\cite{TFIM} with open boundary conditions to examine the performance of TGF and TGFQS. TFIM has been used in the context of quantum annealing~\cite{TFIM_quantum_annealing, TFIM_quantum_annealing_2, TFIM_quantum_annealing_3}, quantum simulation~\cite{TFIM_quantum_simulation}, and VQAs~\cite{TFIM_vqa_sim1, TFIM_vqa_2}. The Hamiltonian of one-dimensional TFIM used in this work is defined as follows

\begin{equation}
    H_{TFIM} = -\mathcal{J} \sum_{i=1}^{n-1} Z_i Z_{i+1} - h \sum_{j=1}^n X_j,
\end{equation}\\
where $\mathcal{J}$ denotes the coupling between the nearest neighbors in a one-dimensional lattice, and $h$ is the transverse magnetic field on the X-axis. We set the parameter values as $\mathcal{J} = h = 0.5$ in this work.

This time, we kept the PQC depth constant while varying the number of qubits. We used 8-, 10-, and 12-qubit TFIM Hamiltonians and set the number of PQC layers to 4. The results for Fraxis, FQS, TGF, and TGFQS of 20 runs over 50 iterations are shown in Fig.~\ref{TFIM_results}. We observe that in most cases, random and half-shifted gate pairing strategies obtain lower relative errors compared to the rest of the optimization methods. Although TGF and TGFQS perform equally well or better than their standard versions, there is a substantial difference in the convergence across gate pairing strategies. In the case of TGF and Fraxis, the linear and opposite gate pairing strategies obtain little to no advantage over standard Fraxis. For TGFQS and standard FQS, the results improve noticeably in most cases compared to TGF. TGF requires three times as many circuit evaluations per updated gate as standard Fraxis, but regardless of the adjusted cost in terms of circuit evaluations, TGF with random and half-shifted gate pairings performs better. The same observation holds for FQS and TGFQS, where TGFQS requires five times more circuit evaluations per gate update. TGFQS generally outperforms the standard FQS when comparisons are made in terms of circuit evaluations. 

In Fig.~\ref{TFIM_results_circuit_evals_8Q_4L}, we compare the two-gate optimization methods with the standard versions, as well as the gradient-based optimization method Adam, with the adjusted circuit evaluations performed in the optimization. For the Adam optimizer, we utilize the parameter-shift rule~\cite{schuld2019evaluating} to compute the gradients. Any single-qubit unitary can be decomposed into a product of three rotation gates~\cite{nielsen2010quantum}

\begin{equation}
     U = e^{i\delta} R_{\hat{n}}(\vartheta) R_{\hat{m}}(\varphi) R_{\hat{n}}(\omega),
\end{equation}\\
for some $\delta, \vartheta, \varphi$ and $\omega$. Here, $\hat{n}$ and $\hat{m}$ are orthogonal rotational axes and $\vartheta, \varphi, \omega$ are rotation angles. We used the $R_Y R_Z$ and $R_Y R_Z R_Y$ decompositions that replace the individual parameterized single-qubit unitaries in the ansatz in Fig.~\ref{ansatz_circuit} and set the learning rate to 0.05 for the Adam optimizer. The reported counts for circuit evaluations include all parameter-shift evaluations required to estimate the gradients of rotation angles in the decomposed ansatz.

Although TGF and TGFQS require substantially more circuit evaluations per gate update, they achieve a lower relative error than the gradient-based Adam and standard Fraxis and FQS. However, this holds only for the best-performing gate pairing strategies, random and half-shifted. With the adjusted circuit evaluations performed in the optimization, the standard Fraxis and FQS have fast initial convergence, but they plateau relatively quickly, whereas TGF and TGFQS continue to converge with random and half-shifted gate pairings. Additionally, despite the additional parameterized gates for Adam, it does not reach as good convergence and relative error as the best performing gate pairings for TGF or TGFQS.

The finite measurement accuracy experiment was repeated for the 8-qubit TFIM Hamiltonian using $L=2$, and shot counts 4096, 8192, and 16384, as in the Fermi--Hubbard case. In addition to random gate pairing, we also examined the performance of the half-shifted gate pairing strategy. The results in Fig.~\ref{TFIM_results_8Q_noisy_2L_noisy} show that TGF retains an advantage over Fraxis in terms of convergence across all tested shot counts. TGFQS, on the other hand, is more sensitive to measurement accuracy. As the shot count increases, its performance improves, indicating its sensitivity to finite measurement accuracy. As in the Fermi--Hubbard case, increasing the number of layers did not improve the performance of TGF and TGFQS relative to Fraxis and FQS, respectively.

\begin{figure}
    \centering
    \includegraphics[width=0.99\linewidth]{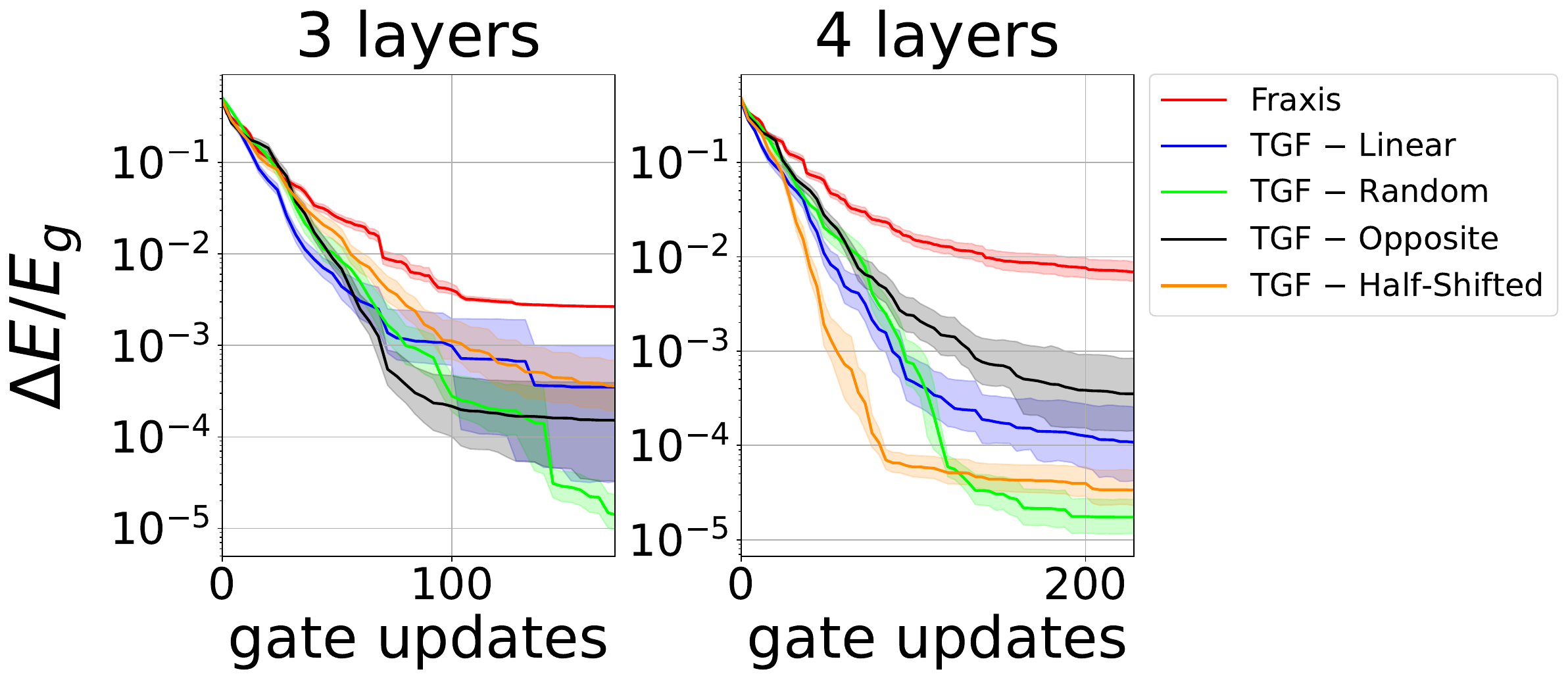}
    \includegraphics[width=0.99\linewidth]{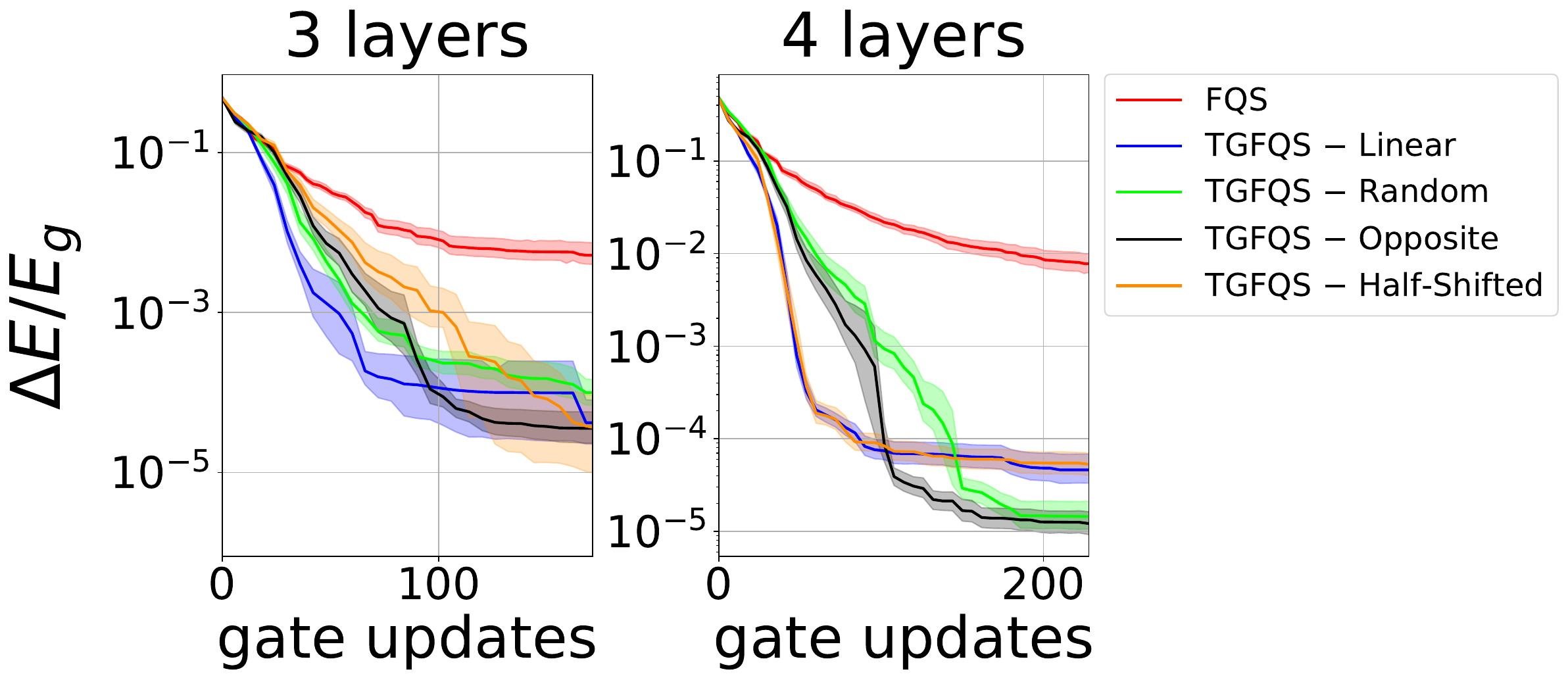}
    \cprotect\caption{Results for 12-qubit LiH molecular Hamiltonian for Fraxis, FQS (red), and two-gate optimizers TGF and TGFQS with linear, random, opposite, and half-shifted gate pairings. The figures are shown on a semi-log scale, with the vertical axis showing the relative error with respect to the ground state energy. Each line indicates a mean of 20 runs, and the shaded areas are 68\% confidence intervals around the mean.}
    \label{LiH_results}
\end{figure}

\begin{figure}
    \centering
    \includegraphics[width=0.99\linewidth]{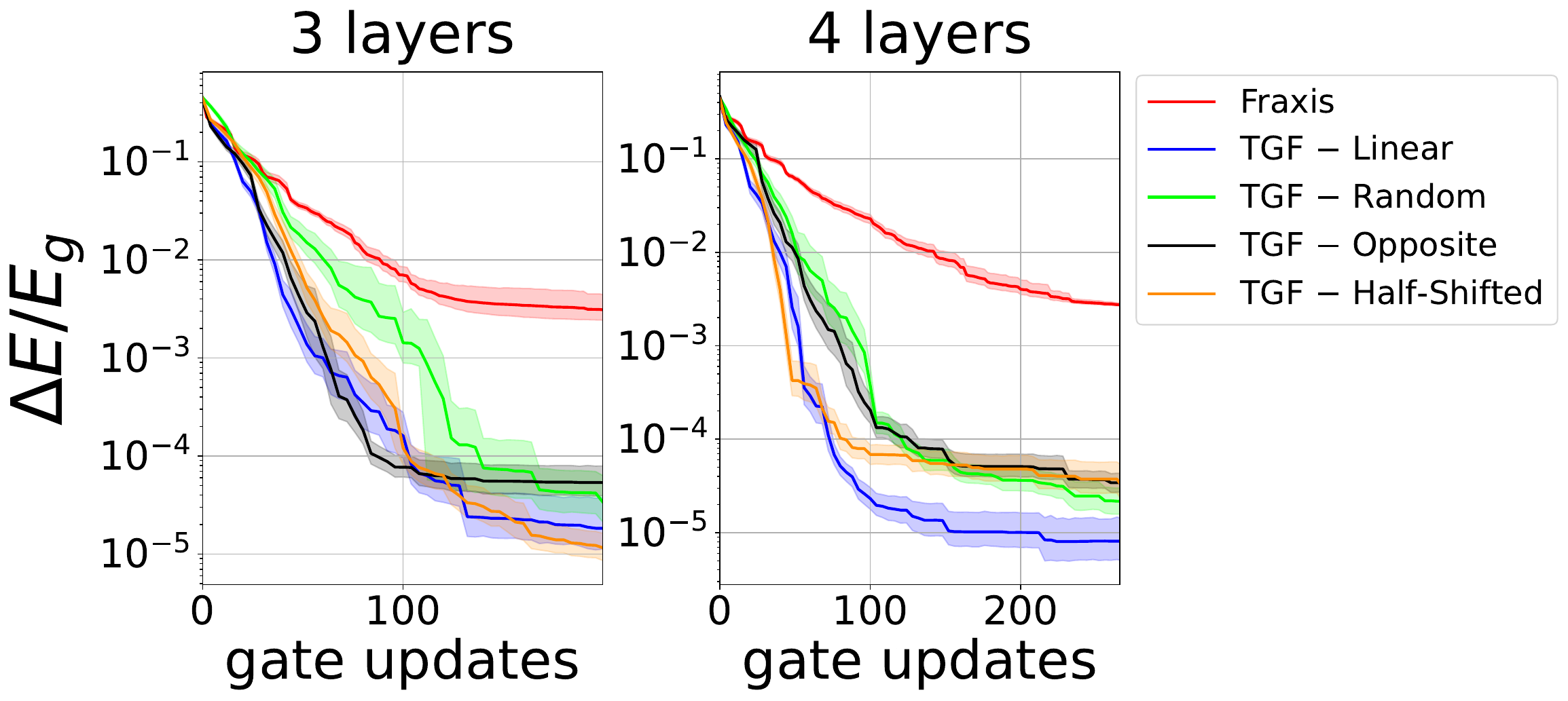}
    \includegraphics[width=0.99\linewidth]{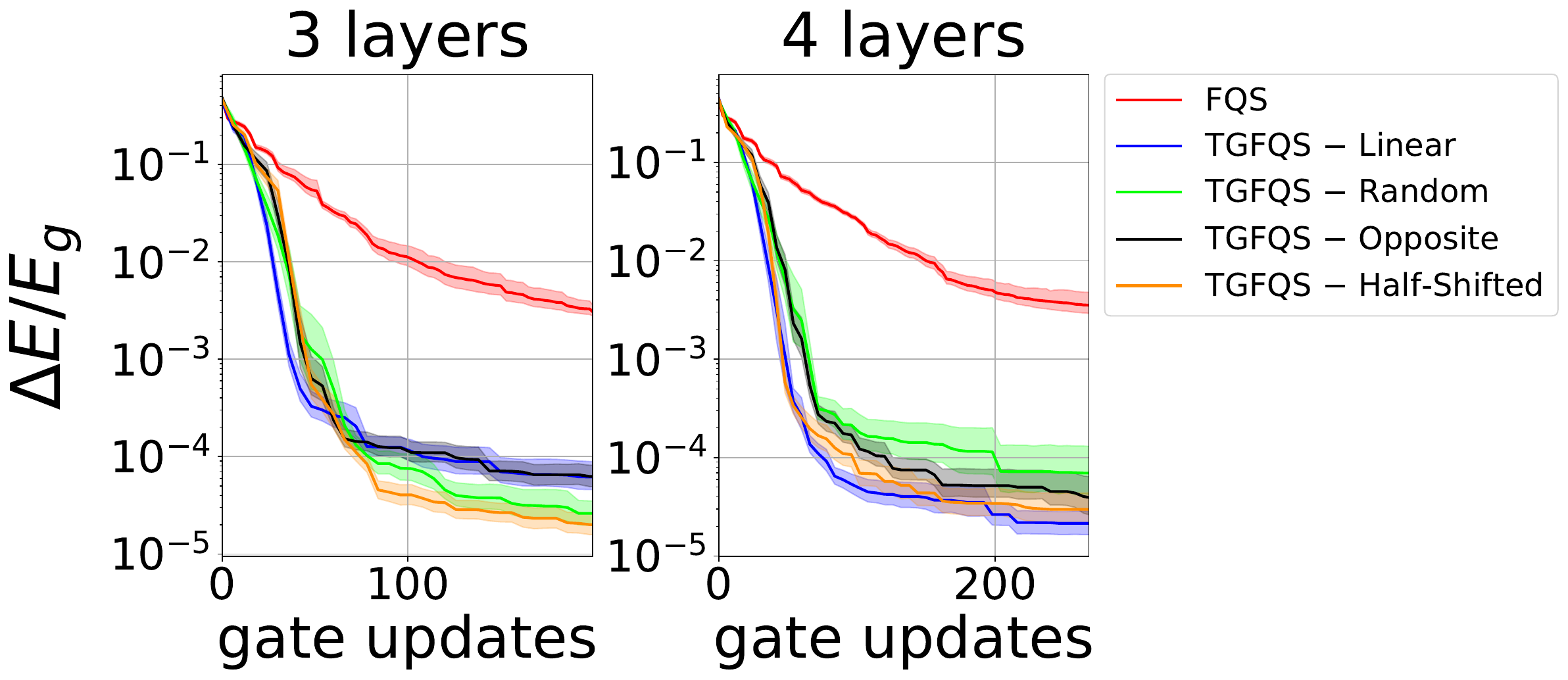}
    \cprotect\caption{Results for 14-qubit BeH$_2$ molecular Hamiltonian for Fraxis, FQS (red), and two-gate optimizers TGF and TGFQS with linear, random, opposite, and half-shifted gate pairings. The figures are shown on a semi-log scale, with the vertical axis showing the relative error with respect to the ground state energy. Each line indicates a mean of 20 runs, and the shaded areas are 68\% confidence intervals around the mean.}
    \label{BeH2_results}
\end{figure}

\subsection{Molecular Hamiltonians}

We use the LiH and BeH$_2$ molecular Hamiltonians as cost functions for ground state optimization with Fraxis, FQS, and their two-gate optimizer counterparts TGF and TGFQS, respectively. The molecular Hamiltonians in the second quantization are given by

\begin{equation}
    H = \sum_{p,q} u_{pq} a_p^\dagger a_q + \frac{1}{2} \sum_{p,q,r,s} u_{pqrs} a_p^\dagger a_q^\dagger a_r a_s,
\end{equation}\\
where $a_p^\dagger$ and $a_q$ are the fermionic creation and annihilation operators. The coefficients $u_{pq}$ and $u_{pqrs}$ represent one- and two-electron integrals, which are written as follows~\cite{one_two_electron_integrals}

\begin{align}
    u_{pq} &= \int dx\psi_p^*(x_1)\left(\frac{\nabla^2}{2} - \frac{Z_I}{|r- R_I|}\right) \psi_q(x_1), \\[0.5cm]
    u_{pqrs} &= \iint dx_1 dx_2 \frac{\psi_p(x_1)^* \psi_q(x_2)^* \psi_r(x_2) \psi_s(x_1)}{r_{12}}.
\end{align}\\
Here, $R_I$ denotes the position of the nuclei in one-electron integral and $r_{12}$ is the distance between electrons in the two-electron integral. Here, $\psi_p(x)$ denotes the $p$-th single-electron orbital.

Finally, we use the Jordan--Wigner mapping~\cite{jordan_wigner} to map the fermionic operators $a_q^\dagger$ and $a_q$ to the Pauli basis. This is done in the following way

\begin{align}
    a_j^\dagger &= \frac{1}{2}(X_j + iY_j) \bigotimes_{r=1}^{j-1} Z_r, \\[0.2cm]
    a_j &= \frac{1}{2}(X_j - iY_j) \bigotimes_{r=1}^{j-1} Z_r.
\end{align}\\
Here, $X_j, Y_j$ and $Z_j$ are the Pauli matrices acting on the $j$-th qubit. We remark that there are other ways to map fermionic operators to the Pauli basis, like Parity~\cite{parity_mapping_ref} and Bravyi--Kitaev~\cite{bravyi_kitaev_mapping}, and other mappings~\cite{other_mappings}. In this work, we use the Jordan--Wigner mapping as it is the most commonly used.

Following this procedure, we mapped the fermionic operators to the Pauli basis and obtained molecular Hamiltonians suitable for PQC optimization. We used the STO-3G basis set~\cite{sto_3g_basis_set_ref} and bond lengths of 1.57 Å and 1.33 Å for LiH and BeH$_2$, respectively. The Hamiltonians for each molecule were extracted from the PennyLane datasets of Refs.~\cite{LiH_Hamiltonian, BeH2_dataset_ref} for LiH and BeH$_2$, respectively. For molecular Hamiltonian experiments, we performed 20 runs for each optimizer, using 5 iterations per run and setting the number of layers to $L=3,4$. The number of iterations was limited to 5, because each iteration is computationally expensive as the number of Pauli strings whose expectation values must be estimated grows rapidly with the number of qubits~\cite{VQE_practices_and_methods}.

\begin{figure}
    \centering
    \includegraphics[width=0.99\linewidth]{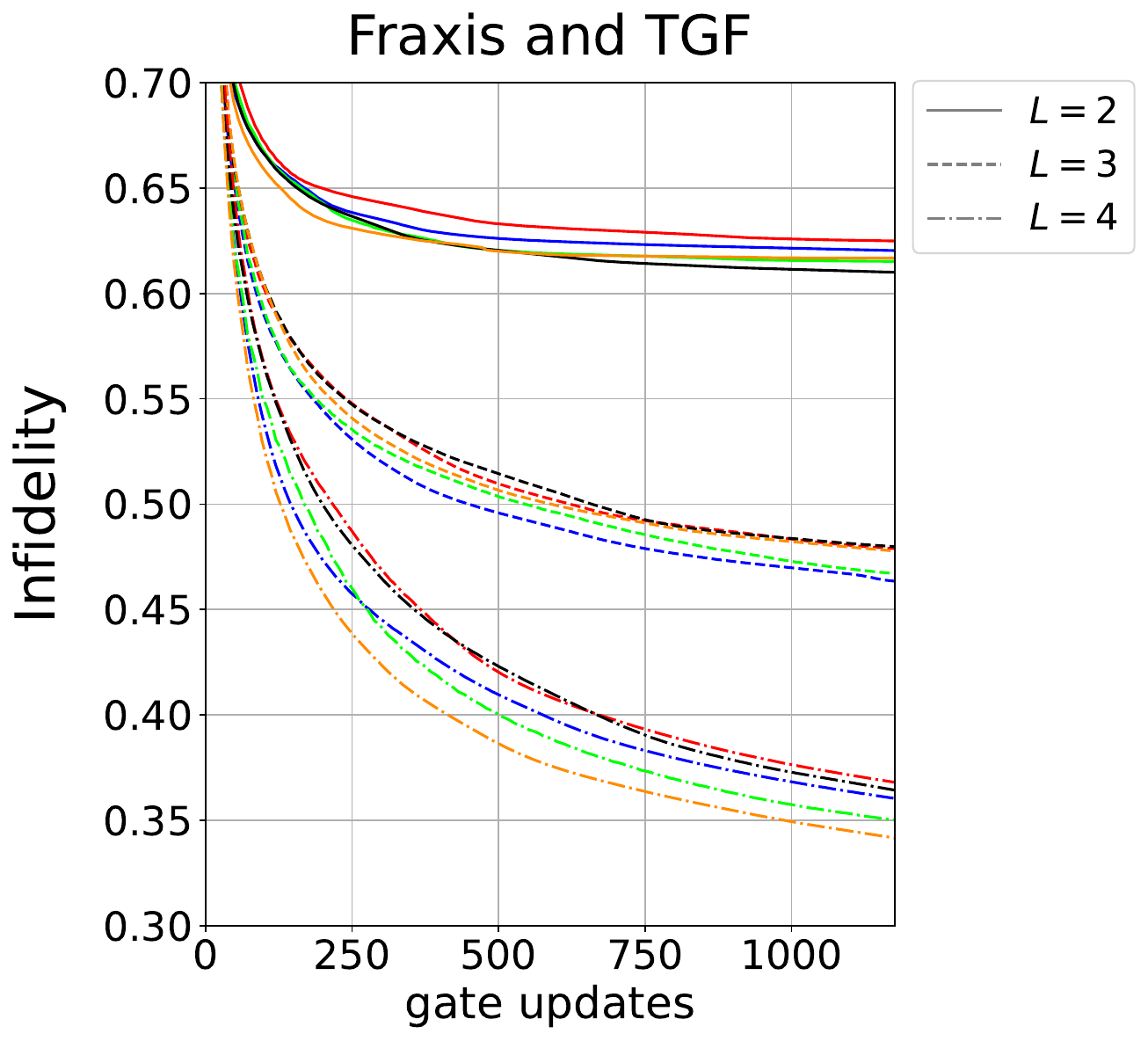}
    \\[0.5cm]
    \includegraphics[width=0.99\linewidth]{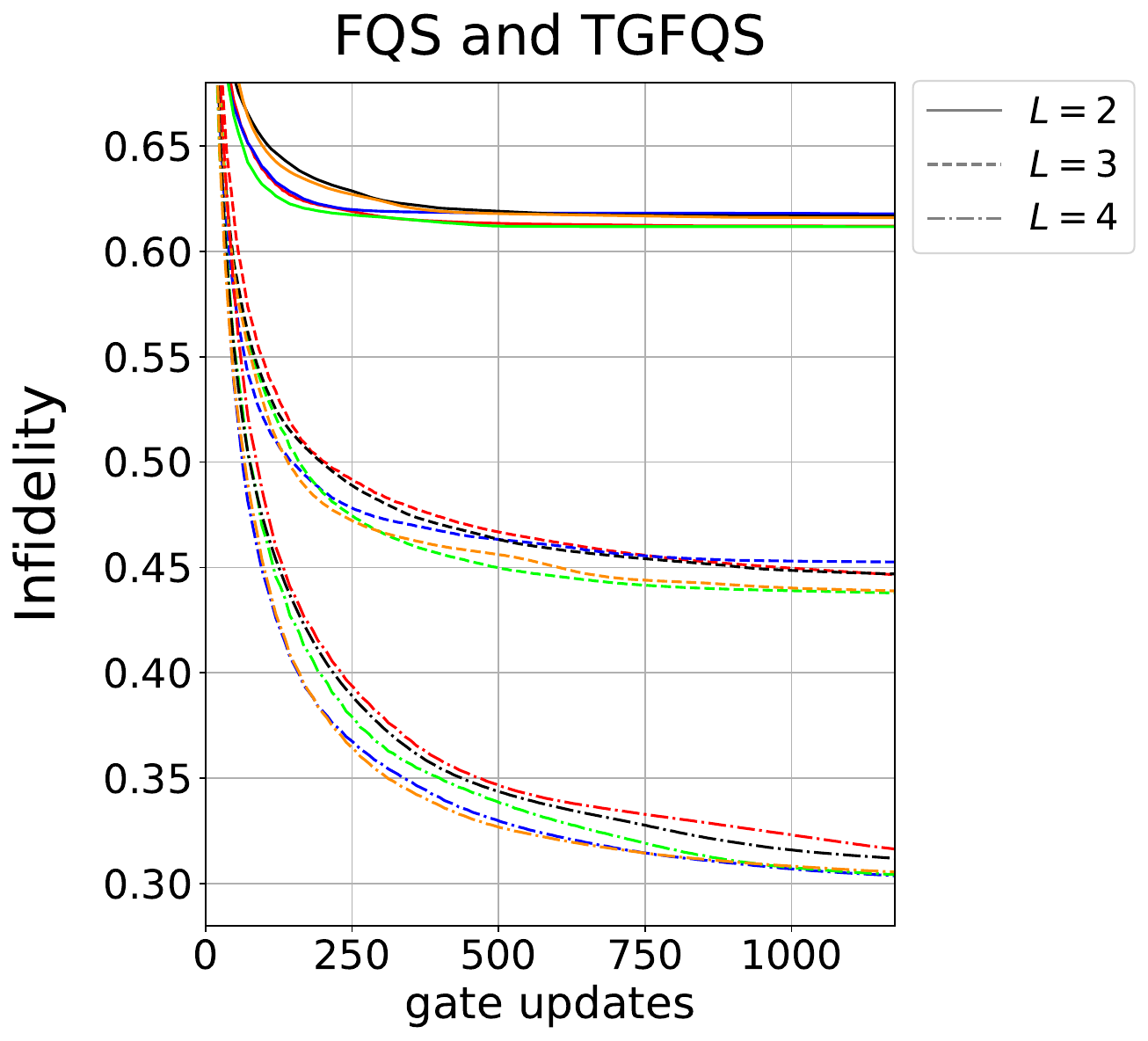}
    \cprotect\caption{Minimizing the infidelity using the single-gate optimizers Fraxis, FQS (red), and two-gate optimizers TGF and TGFQS with the following gate pairings: linear (blue), random (green), opposite (black), and half-shifted (orange). Each line represents the mean of 50 runs using 6 qubits with $L = 2$ (solid line), $L=3$ (dashed line), and $L=4$ (dash-dotted line) layers.}
    \label{random_state_results}
\end{figure}

We present the results for LiH and BeH$_2$ Hamiltonians in Figs.~\ref{LiH_results} and~\ref{BeH2_results}, respectively. In all figures, we observe that the two-gate optimizers TGF and TGFQS achieve lower relative error than their single-gate optimizer counterparts. Despite the increased cost per gate in terms of circuit evaluations, the two-gate optimizers achieve lower relative error than Fraxis and FQS in both molecular Hamiltonian benchmarks. Since the circuit evaluations required to determine the coefficients $\rr$ in Eq.~(\ref{tgfqs_cost_function}) are independent, they could in principle be executed in parallel using multiple quantum devices. This may reduce the overhead caused by the large number of circuit evaluations to construct the local two-gate cost functions.

\subsection{Fidelity Maximization}

In the final experiment, we maximize the fidelity between a randomly sampled target state $\ket{\phi}$ and the state prepared by the PQC, $\ket{\Psi(\tet)}$. Fidelity is a commonly used metric to measure the closeness between two quantum states~\cite{nielsen2010quantum, wilde_quantum_information_book} and has been used in quantum state preparation~\cite{quantum_state_prep_1, quantum_state_prep_magic, quantum_state_prep_3}, quantum kernel methods~\cite{QML_fidelity_kernel, QML_fidelity_kernel_benchmark}, and benchmarking quantum hardware~\cite{quantum_hardware_benchmark_1, quantum_hardware_benchmark_2, quantum_hardware_benchmark_3}. For two pure quantum states $\ket{\varphi}$ and $\ket{\chi}$ the fidelity $F(\ket\varphi, \ket{\chi})$ is defined as follows

\begin{equation}
    F(\ket\varphi, \ket{\chi}) = |\langle \varphi | \chi \rangle|^2.
\end{equation}  

In this work, we minimize infidelity, which is equivalent to maximizing fidelity. Thus, the cost function we use in this section is defined as 

\begin{equation}
    \text{infidelity} = 1 - |\langle \phi |\Psi(\tet)\rangle|^2,
\end{equation}\\
where $\ket{\phi}$ is the random target state and $\ket{\Psi(\tet)}$ is the state produced by the PQC. We carried out a total of 50 runs for all optimizers. At the beginning of each run, we sampled a new random target state $\ket{\phi}$ to maximize the fidelity with.

In Fig.~\ref{random_state_results}, we display the results for fidelity maximization with 6 qubits while setting the number of layers to $L=2,3,4$. The standard versions of Fraxis and FQS are colored red, whereas the two-gate methods with pairings are colored as follows: linear, blue; random, green; opposite, black; half-shifted, orange. The linestyle indicates the number of layers that were used for the experiment: solid, dashed, and dash-dotted lines correspond to $L=2$, $L=3$, and $L=4$, respectively. As the number of layers increases in the system, the performance gap between single- and two-gate optimizers widens, and the best-performing gate pairing strategies achieve lower final infidelity than standard Fraxis and FQS. This only holds for random and half-shifted gate pairing strategies, which perform the best. This suggests that for the quantum state preparation, the two-gate local optimization can be advantageous when combined with an effective gate pairing strategy.

\begin{table*}
\centering
\small
\setlength{\tabcolsep}{4pt}
\renewcommand{\arraystretch}{1.25}
\begin{tabular}{|l|c|c|c|c|c|c|}
\hline
\multirow{2}{*}{Cost Function} 
& \multicolumn{2}{c|}{Baseline $(\Delta E / E_g)$} 
& \multicolumn{2}{c|}{Best relative improvement (\%)} 
& \multicolumn{2}{c|}{Best gate pairing strategy} \\
\cline{2-7}
& Fraxis & FQS & TGF & TGFQS & TGF & TGFQS \\
\hline
\hline
Fermi--Hubbard
& $4.8 \cdot 10^{-2}$ & $8.8 \cdot 10^{-4}$ &  \hspace{0.5cm} 99.0\% \hspace{0.5cm} & 96.7\% 
& Random & Random \\
\hline
TFIM (12 qubits)
& $2.1 \cdot 10^{-2}$ & $2.1 \cdot 10^{-2}$ & 92.0\% & 99.98\%
& Half-Shifted & Half-Shifted \\
\hline
LiH $(L=4)$ 
& $6.4 \cdot 10^{-3}$ & $7.4 \cdot 10^{-3}$ & 99.7\% & 99.8\%
& Random & Opposite \\
\hline
BeH$_2$ $(L=4)$ 
& $2.7 \cdot 10^{-3}$ & $3.5 \cdot 10^{-3}$ & 99.8\% & 99.4\%
& Linear & Linear \\
\hline
Fidelity maximization $(L=4)$
& 0.367 & 0.316 & 7.2\% & 3.9\% 
& Half-Shifted & Random \\
\hline
\end{tabular}
\caption{Summary of benchmark tasks, the baseline final relative errors for Fraxis and FQS, as well as the best-performing gate pairing strategy for TGF and TGFQS with the best relative improvement. For the fidelity maximization task, the reported quantity is the final infidelity at the end of optimization.}
\label{summary_table}
\end{table*}

\section{Conclusions} \label{conclusion_section}

In this work, we introduced two-gate extensions of the sequential optimizers Fraxis and FQS, termed TGF and TGFQS, respectively. These methods simultaneously optimize an arbitrary pair of parameterized single-qubit gates in the circuit, therefore enabling different gate pairing strategies to be used in the optimization. We benchmarked the proposed methods using the Fermi--Hubbard and transverse-field Ising model spin Hamiltonians, first comparing the convergence in terms of gate updates. We then compared TGF and TGFQS against the gradient-based optimizer Adam using $R_Y R_Z$ and $R_Y R_Z R_Y$ decompositions and evaluated the convergence as a function of circuit evaluations. We also compared TGF and TGFQS with the best-performing gate pairing strategies with finite measurement accuracy using 4096, 8192, and 16384 shots to approximate each Hamiltonian term. We then examined performance for the 12-qubit LiH and 14-qubit BeH$_2$ molecular Hamiltonians, and for a 6-qubit quantum state preparation task based on fidelity maximization. 

The results are summarized in Table~\ref{summary_table}, which reports the mean final value across the runs for Fraxis and FQS. The table also lists the best-performing gate pairing strategies for TGF and TGFQS, with the best relative improvement compared to the baseline for each benchmark task under the noiseless statevector simulation. The improved performance of TGF and TGFQS over Fraxis and FQS comes at the cost of additional circuit evaluations: TGF and TGFQS require a total of 36 and 100 circuit evaluations per gate pair update, corresponding to 18 and 50 circuit evaluations per updated gate, respectively. Fraxis and FQS, on the other hand, require only 6 and 10 circuit evaluations per gate update, respectively. Empirically, the half-shifted and random gate pairing strategies performed best in most settings, whereas the opposite and linear gate pairings yielded only modest improvements over the baseline methods. 

The circuit evaluations required to construct the local two-gate cost functions are mutually independent and therefore suitable for parallel execution across multiple quantum devices. Consequently, the practical overhead of TGF and TGFQS may be reduced in settings where circuit evaluations can be executed in parallel. A natural direction for further work is to extend the method to the simultaneous optimization of more than two gates. However, this becomes a challenging task as the required measurements grow as $6^m$ when optimizing $m$ gates simultaneously in the Fraxis-based local cost function, and as $10^m$ in the FQS-based local cost function. In addition, the number of distinct terms $\A_{\mu \nu \alpha \beta}$ appearing in the cost function grows exponentially. For simultaneous optimization of $m$ parameterized single-qubit gates, the number of such terms scales as $9^m$ for Fraxis and $16^m$ for FQS. Future work could also investigate approximation strategies for the simultaneous multi-gate optimization that reduce the number of required circuit evaluations or the number of terms in the cost function, with a controlled loss of accuracy. 

\section{Data Availability}

The data used to generate the figures in this work are available upon reasonable request. 

\begin{acknowledgments}
The author acknowledges funding by Business Finland for the project 8726/31/2022 CICAQU. The author received funding from InstituteQ's doctoral school. 
\end{acknowledgments}

\appendix

\section{Derivation of Two-Gate Coefficient Formulas} \label{R_2_and_R_4_trace_calculation_appendix}

In this appendix, we derive the formulas for the coefficients $\rr$ in Eqs.~(\ref{R2_term_equation})--(\ref{R4_term_equation}) of the main text. We begin by recalling the definitions of the coefficients for $\rr$, which were given as
\begin{align}
    &\rr\left( \varsigma_{d,\mu}', \varsigma_{k,\alpha}' \right) \coloneqq \A_{\mu \mu \alpha \alpha},\\[0.2cm]
    &\rr\left( \varsigma_{d,\mu}', \varsigma_{k,\alpha}', \varsigma_{k,\beta}' \right) \coloneqq \A_{\mu \mu \alpha \beta} + \A_{\mu \mu \beta \alpha}, \quad \alpha \neq \beta, \\[0.2cm]
    &\rr\left( \varsigma_{d,\mu}', \varsigma_{d,\nu}', \varsigma_{k,\alpha}' \right) \coloneqq \A_{\mu \nu \alpha \alpha} + \A_{\nu \mu \alpha \alpha}, \quad \mu \neq \nu,  \\[0.2cm]
    &\rr\left( \varsigma_{d,\mu}', \varsigma_{d,\nu}', \varsigma_{k,\alpha}', \varsigma_{k,\beta}' \right)  \coloneqq \A_{\mu \nu \alpha \beta} +  \A_{\mu \nu \beta \alpha} + \A_{\nu \mu \alpha \beta} \nonumber \\ 
    & \hspace{3.5cm} + \A_{\nu \mu \beta \alpha}, \quad  \mu \neq \nu, \ \alpha \neq \beta,
\end{align}
where

\begin{equation}
    \A_{\mu \nu \alpha \beta} \coloneqq \Tr \left(M'\varsigma_{d,\mu}' V \varsigma_{k,\alpha}' \rho' {\varsigma_{k,\beta}'}^\dagger V^\dagger {\varsigma_{d,\nu}'}^\dagger \right).
\end{equation}

For notational convenience, we define
\begin{equation}
    \trt\left(O_d, O_k \right) \coloneqq \Tr(M' O_d V O_k \rho' O_k^\dagger V^\dagger O_d ^\dagger),
\end{equation}
where $O_d$ and $O_k$ denote the operators that are inserted in place of the $d$-th and $k$-th parameterized gates, respectively. The coefficients $\rr\left( \varsigma_{d,\mu}', \varsigma_{k,\alpha}' \right)$ are obtained directly as
\begin{equation}\label{R_1_coefficient_eq}
    \rr\left( \varsigma_{d,\mu}', \varsigma_{k,\alpha}' \right) = \trt\left( \varsigma_{d,\mu}', \varsigma_{k,\alpha}' \right).
\end{equation}

To derive the cubic coefficients of $\rr$, we expand $\trt\left( \varsigma_{d, \mu}', \varsigma_{k, (\alpha + \beta)}' \right)$ in terms of $\A_{\mu \nu \alpha \beta}$ as follows

\begin{widetext}
\begin{align}\label{R_2_equation_A_index}
    \trt\left( \varsigma_{d, \mu}', \varsigma_{k, (\alpha + \beta)}' \right) = \Tr\Biggl[M' \varsigma_{d, \mu}' V \left(\frac{\varsigma_{k, \alpha}' + \varsigma_{k, \beta}'}{\sqrt{2}} \right) \rho' \left(\frac{\varsigma_{k, \alpha}' + \varsigma_{k, \beta}'}{\sqrt{2}} \right)^\dagger V^\dagger  {\varsigma_{d, \mu}'}^\dagger \Biggl] &= \frac{1}{2} (\A_{\mu \mu \alpha \alpha} + \A_{\mu \mu \beta \beta} + \A_{\mu \mu \alpha \beta} + \A_{\mu \mu \beta \alpha}),
\end{align}
\end{widetext}
where $\alpha \neq \beta$. Here, $\A_{\mu \mu \alpha \alpha} = \trt\left( \varsigma_{d, \mu}', \varsigma_{k, \alpha}' \right)$ and $\A_{\mu \mu \beta \beta} = \trt\left( \varsigma_{d, \mu}', \varsigma_{k, \beta}' \right)$. The left-hand side of Eq.~\ref{R_2_equation_A_index} is $\trt\left( \varsigma_{d, \mu}', \varsigma_{k, (\alpha + \beta)}' \right)$ and therefore we obtain

\begin{align}\label{R_2_eq_final_derivation}
\begin{split}
    &\rr\left( \varsigma_{d,\mu}', \varsigma_{k,\alpha}', \varsigma_{k,\beta}' \right) = \A_{\mu \mu \alpha \beta} + \A_{\mu \mu \beta \alpha} \\[0.1cm]
    & =2\cdot \trt\left( \varsigma_{d, \mu}', \varsigma_{k, (\alpha + \beta)}' \right) -  \trt\left( \varsigma_{d, \mu}', \varsigma_{k, \alpha}' \right) - \trt\left( \varsigma_{d, \mu}', \varsigma_{k, \beta}' \right),
\end{split}
\end{align}
where $\alpha \neq \beta$. An analogous calculation yields the corresponding expression for $\rr\left( \varsigma_{d, \mu}', \varsigma_{d, \nu}', \varsigma_{k, \alpha}' \right)$. 

To derive the quartic coefficients $\rr\left( \varsigma_{d, \mu}', \varsigma_{d, \nu}', \varsigma_{k, \alpha}', \varsigma_{k, \beta}' \right)$, we expand $\trt\left( \varsigma_{d, (\mu+\nu)}', \varsigma_{k, (\alpha + \beta)}' \right)$ in terms of $\A_{\mu \nu \alpha \beta}$

\begin{widetext}
\begin{align}\label{R_4_trace_calc_equation_last}
\begin{split}
    \trt\left( \varsigma_{d, (\mu+\nu)}', \varsigma_{k, (\alpha + \beta)}' \right) &= \Tr\Biggl[M' \left(\frac{\varsigma_{d, \mu}' + \varsigma_{d, \nu}'}{\sqrt{2}} \right) V \left(\frac{\varsigma_{k, \alpha}' + \varsigma_{k, \beta}'}{\sqrt{2}} \right) \rho' \left(\frac{\varsigma_{k, \alpha}' + \varsigma_{k, \beta}'}{\sqrt{2}} \right)^\dagger V^\dagger  \left(\frac{\varsigma_{d, \mu}' + \varsigma_{d, \nu}'}{\sqrt{2}} \right)^\dagger \Biggl]  \\[0.2cm]
     & = \frac{1}{4} \Bigl[ \A_{\mu \mu \alpha \alpha} + \A_{\mu \mu \beta \beta} + \A_{\nu \nu \alpha \alpha} + \A_{\nu \nu \beta \beta} \\[0.2cm]
     & \quad + (\A_{\mu \mu \alpha \beta} + \A_{\mu \mu \beta \alpha}) + (\A_{\nu \nu \alpha \beta} + \A_{\nu \nu \beta \alpha}) + (\A_{\mu \nu \alpha \alpha} + \A_{\nu \mu \alpha \alpha}) +  (\A_{\mu \nu \beta \beta} + \A_{\nu \mu \beta \beta}) \\[0.2cm]
     & \quad + (\A_{\mu \nu \alpha \beta} + \A_{\mu \nu \beta \alpha} + \A_{\nu \mu \alpha \beta} + \A_{\nu \mu \beta \alpha})\Bigl],
\end{split}
\end{align}
\end{widetext}
where $\alpha \neq \beta$ and $\mu \neq \nu$.  Here, the terms are grouped so that the contributions already expressible through expectation values $\trt(\cdot, \cdot)$ are separated from the remaining quartic coefficient. $\A_{\mu \mu \alpha \alpha}$ and the terms $(\A_{\mu \mu \alpha \beta} + \A_{\mu \mu \beta \alpha})$ can be computed using Eqs.~(\ref{R_1_coefficient_eq}) and~(\ref{R_2_eq_final_derivation}), respectively, and the final row of Eq.~(\ref{R_4_trace_calc_equation_last}) represents the coefficients $\rr\left( \varsigma_{d, \mu}', \varsigma_{d, \nu}', \varsigma_{k, \alpha}', \varsigma_{k, \beta}' \right)$. Isolating the remaining quartic term then yields 

\begin{widetext}
\begin{align}
\begin{split}
    \rr \bigl( \varsigma_{d,\mu}',& \varsigma_{d,\nu}', \varsigma_{k,\alpha}', \varsigma_{k,\beta}' \bigr) =  \A_{\mu \nu \alpha \beta} + \A_{\mu \nu \beta \alpha} + \A_{\nu \mu \alpha \beta} + \A_{\nu \mu \beta \alpha} \\[0.2cm]
   &= 4 \cdot \trt\left(\varsigma_{d, (\mu+\nu)}', \varsigma_{k, (\alpha + \beta)}'\right) -  \trt\left(\varsigma_{d, \mu}', \varsigma_{k, \alpha}'\right) -  \trt\left(\varsigma_{d, \mu}', \varsigma_{k, \beta}'\right) -  \trt\left(\varsigma_{d, \nu}', \varsigma_{k, \alpha}'\right) -  \trt \left(\varsigma_{d, \nu}', \varsigma_{k, \beta}' \right) \\[0.2cm]
& \quad- \Bigl[ 2\cdot \trt \left(\varsigma_{d, \mu}', \varsigma_{k, (\alpha + \beta)}' \right) -  \trt \left(\varsigma_{d, \mu}', \varsigma_{k, \alpha}' \right) - \trt \left(\varsigma_{d, \mu}', \varsigma_{k, \beta}' \right)\Bigr] - \Bigl[ 2\cdot \trt \left(\varsigma_{d, \nu}', \varsigma_{k, (\alpha + \beta)}' \right) -  \trt \left(\varsigma_{d, \nu}', \varsigma_{k, \alpha}' \right) - \trt \left(\varsigma_{d, \nu}', \varsigma_{k, \beta}' \right) \Bigr] \\[0.2cm]
& \quad - \Bigl[ 2\cdot \trt \left(\varsigma_{d, (\mu + \nu)}', \varsigma_{k, \alpha}' \right) -  \trt \left(\varsigma_{d, \mu}', \varsigma_{k, \alpha}' \right) - \trt \left(\varsigma_{d, \nu}', \varsigma_{k, \alpha}' \right)\Bigr] - \Bigl[ 2\cdot \trt \left(\varsigma_{d, (\mu + \nu)}', \varsigma_{k, \beta}' \right) -  \trt \left(\varsigma_{d, \mu}', \varsigma_{k, \beta}' \right) - \trt \left(\varsigma_{d, \nu}', \varsigma_{k, \beta}' \right) \Bigr] \\[0.4cm]
& = 4 \cdot \trt \left(\varsigma_{d, (\mu+\nu)}', \varsigma_{k, (\alpha + \beta)}' \right)  +  \trt \left(\varsigma_{d, \mu}', \varsigma_{k, \alpha}' \right) +  \trt \left(\varsigma_{d, \mu}', \varsigma_{k, \beta}' \right) +  \trt \left(\varsigma_{d, \nu}', \varsigma_{k, \alpha}' \right) +  \trt \left(\varsigma_{d, \nu}', \varsigma_{k, \beta}' \right) \\[0.2cm]
& \quad - 2\cdot \trt \left(\varsigma_{d, \mu}', \varsigma_{k, (\alpha + \beta)}' \right) - 2\cdot \trt \left(\varsigma_{d, \nu}', \varsigma_{k, (\alpha + \beta)}' \right) - 2\cdot \trt \left(\varsigma_{d, (\mu + \nu)}', \varsigma_{k, \alpha}' \right) -  2\cdot \trt \left(\varsigma_{d, (\mu + \nu)}', \varsigma_{k, \beta}' \right).
\end{split}
\end{align}
\end{widetext}
This completes the derivation of the coefficients $\rr$ used in Eqs.~(\ref{R2_term_equation})--(\ref{R4_term_equation}) of the main text.

\bibliography{refs}
\end{document}